
\def\ignore#1{}
 

\newcount\sectnum
\newcount\subsectnum
\newcount\eqnumber

\global\eqnumber=1\sectnum=0


\def\lab{(\the\sectnum.\the\eqnumber)}



\def\show#1{#1}



\def\smskip{\vskip 5 pt}
\def\medskip{\vskip 10 pt}
\def\bigskip{\vskip 15 pt}
\def\pn{\par\noindent}
\def\br{\break}

\def\bl{\bigl} 
\def\br{\bigr} 
\def\lf{\left}
\def\ri{\right}

\def\ol#1{\overline{#1}}

\def\a{\alpha}

\def\b{\beta}
\def\l{\lambda}
\def\g{\gamma}
\def\m{\mu}

\def\d{\delta}

\def\P{\Pi}

\def\re{\Re}
\def\rn{\Re^n}

\def\tl{\tilde}

\def\old#1{}
\def\leaderfill{\leaders\hbox to 1em{\hss.\hss}\hfill}


\parindent=2pc
\baselineskip=15pt
\vsize=8.7 true in
\voffset=0.125 true in
\parskip=3pt


\def\minprob#1#2#3{$$\eqalign{&\hbox{minimize\ \ }#1\cr &\hbox{subject to\ \
}#2\cr}\ifnum 0=#3{}\else\eqno(#3)\fi$$}        
     
\def\maxprob#1#2#3{$$\eqalign{&\hbox{maximize\ \ }#1\cr &\hbox{subject to\ \
}#2\cr}\ifnum 0=#3{}\else\eqno(#3)\fi$$}        
     
\def\aligntwo#1#2#3#4#5{$$\eqalign{#1&#2\cr #3&#4\cr}
\ifnum 0=#5{}\else\eqno(#5)\fi$$}
\def\alignthree#1#2#3#4#5#6#7{$$\eqalign{#1&#2\cr #3&#4\cr #5&#6\cr}
\ifnum 0=#7{}\else\eqno(#7)\fi$$}


\def\eqnum{\eqno{\hbox{(\the\sectnum.\the\eqnumber)}\global\advance\eqnumber
by1}}

\def\eqnu{\eqno{\hbox{(\the\sectnum.\the\eqnumber)}\global\advance\eqnumber
by1}}

\newcount\examplnumber
\def\examplnum{\global\advance\examplnumber by1}

\newcount\figrnumber
\def\figrnum{\global\advance\figrnumber by1}

\newcount\propnumber
\def\propnum{\global\advance\propnumber by1}

\newcount\defnumber
\def\defnum{\global\advance\defnumber by1}

\newcount\lemmanumber
\def\lemmanum{\global\advance\lemmanumber by1}

\newcount\assumptionnumber
\def\assumptionnum{\global\advance\assumptionnumber by1}

\def\exampl{\the\sectnum.\the\examplnumber}
\def\figr{\the\sectnum.\the\figrnumber}
\def\propn{\the\sectnum.\the\propnumber}
\def\defn{\the\sectnum.\the\defnumber}
\def\lemman{\the\sectnum.\the\lemmanumber}
\def\assumptionn{\the\sectnum.\the\assumptionnumber}

\def\section#1{\goodbreak\vskip 3pc plus 6pt minus 3pt\leftskip=-2pc
   \global\advance\sectnum by 1\eqnumber=1
\global\examplnumber=1\figrnumber=1\propnumber=1\defnumber=1\lemmanumber=1\assumptionnumber=1%
   \line{\hfuzz=1pc{\hbox to 3pc{\bf 
   \vtop{\hfuzz=1pc\hsize=38pc\hyphenpenalty=10000\noindent\uppercase{\the\sectnum.\quad #1}}\hss}}
			\hfill}
			\leftskip=0pc\nobreak\tenf
			\vskip 1pc plus 4pt minus 2pt\noindent\ignorespaces}



\def\sect#1{\noindent\leftskip=-2pc\tenf
   \goodbreak\vskip 1pc plus 4pt minus 2pt
                \global\advance\subsectnum by 1\eqnumber=1
   \line{\hfuzz=1pc{\hbox to 3pc{\bf 
   \vtop{\hfuzz=1pc\hsize=38pc\hyphenpenalty=10000\noindent\uppercase{{\bf #1}}}\hss}}
                        \hfill}
   \leftskip=0pc\nobreak\tenf
                        \vskip 1pc plus 4pt minus 2pt\nobreak\noindent\ignorespaces}

\def\subsection#1{\noindent\leftskip=0pc\tenf
   \goodbreak\vskip 1pc plus 4pt minus 2pt
   \line{\hfuzz=1pc{\hbox to 3pc{\bf 
   \vtop{\hfuzz=1pc\hsize=38pc\hyphenpenalty=10000\noindent{\bf #1}}\hss}}
                        \hfill}
   \leftskip=0pc\nobreak\tenf
                        \vskip 1pc plus 4pt minus 2pt\nobreak\noindent\ignorespaces}
\def\subsubsection#1{\goodbreak\vskip 1pc plus 4pt minus 2pt
   \hfuzz=3pc\leftskip=0pc\noindent\tenit #1 \nobreak\tenf\vskip 6pt plus 1pt
                                minus 1pt\nobreak\ignorespaces\leftskip=0pc}
%

\def\beginexample#1{\noindent\goodbreak\vskip 6pt plus 1pt minus 1pt
\noindent
  \hbox {\bf Example #1\hss}
  \nobreak\vskip 4pt plus 1pt minus 1pt \nobreak\noindent\ninef
  \global\advance
                \leftskip by\parindent\pn}
\def\endexample{\vskip 12pt\tenf\par
  \global\advance\leftskip by -\parindent
  }

\def\beginexercise#1{\noindent\goodbreak\vskip 6pt plus 1pt minus 1pt \noindent\global\normalbaselineskip=12pt
  \hbox {\bf Exercise #1\hss}
  \nobreak\vskip 4pt plus 1pt minus 1pt 
  \nobreak\noindent\ninef\global\advance\leftskip
                        by\parindent\pn}
\def\endexercise{\vskip 12pt\tenf\par
  \global\advance\leftskip by -\parindent
  }

\def\beginsection#1{\noindent\goodbreak\vskip 6pt plus 1pt minus 1pt \noindent\global\normalbaselineskip=12pt
  \hbox {\it #1\hss}
  \vskip 0.1pt plus 1pt minus 1pt \nobreak\noindent\ninef\global\advance
                \leftskip by\parindent\noindent\pn}
\def\endsection{\vskip 12pt\tenf\par
  \global\advance\leftskip by -\parindent
}

%


\def\section#1{\goodbreak\vskip 3pc plus 6pt minus 3pt\leftskip=-2pc
   \global\advance\sectnum by 1\eqnumber=1
\global\examplnumber=1\figrnumber=1\propnumber=1\defnumber=1\lemmanumber=1\assumptionnumber=1\subsectnum=0%
   \line{\hfuzz=1pc{\hbox to 3pc{\bf 
   \vtop{\hfuzz=1pc\hsize=38pc\hyphenpenalty=10000\noindent\uppercase{\the\sectnum.\quad #1}}\hss}}
			\hfill}
			\leftskip=0pc\nobreak\tenf
			\vskip 1pc plus 4pt minus 2pt\noindent\ignorespaces}

\def\subsection#1{\noindent\leftskip=0pc\tenf
   \goodbreak\vskip 1pc plus 4pt minus 2pt
               \global\advance\subsectnum by 1
   \line{\hfuzz=1pc{\hbox to 3pc{\bf \the\sectnum.\the\subsectnum\ \ \
   \vtop{\hfuzz=1pc\hsize=38pc\hyphenpenalty=10000\noindent{\bf #1}}\hss}}
                        \hfill}
   \leftskip=0pc\nobreak\tenf
                        \vskip 1pc plus 4pt minus 2pt\nobreak\noindent\ignorespaces}

\def\subsubsection#1{\goodbreak\vskip 1pc plus 4pt minus 2pt
   \hfuzz=3pc\leftskip=0pc\noindent{\bf #1} \nobreak\vskip 6pt plus 1pt
                                minus 1pt\nobreak\ignorespaces\leftskip=0pc}



\def\proposition#1{\smskip\pn{\bf Proposition #1}\quad}
\def\proof{\smskip\pn{\bf Proof:}\quad}

 \def\qed{\quad{\bf
Q.E.D.} \par\bigskip}
\def\ref{\smskip\pn}

\def\chapter#1#2{{\bf \centerline{\helbigbig
{#1}}}\bigskip\bigskip{\bf \centerline{\helbigbig
{#2}}}\bigskip\bigskip} 



\def\longpapertitle#1#2#3{{\bf \centerline{\helbigb
{#1}}}\bigskip{\bf \centerline{\helbigb
{#2}}}\bigskip\bigskip{\centerline{
by}}\bigskip{\bf \centerline{
{#3}}}\bigskip\bigskip} 


\def\nitem#1{\smskip\item{#1}}

\newcount\alphanum
\newcount\romnum

\def\alphaenumerate{\ifcase\alphanum \or (a)\or (b)\or (c)\or (d)\or (e)\or
(f)\or (g)\or (h)\or (i)\or (j)\or (k)\fi}
\def\romenumerate{\ifcase\romnum \or (i)\or (ii)\or (iii)\or (iv)\or (v)\or
(vi)\or (vii)\or (viii)\or (ix)\or (x)\or (xi)\fi}

\def\alist{\begingroup\vskip10pt\alphanum=1
\parskip=2pt\parindent=0pt \leftskip=3pc
\everypar{\llap{{\rm\alphaenumerate\hskip1em}}\advance\alphanum by1}}

\def\nolist{\begingroup\vskip10pt\alphanum=0
\parskip=2pt\parindent=0pt \leftskip=3pc
\everypar{\llap{\global\advance\alphanum by1(\the\alphanum)\hskip1em}}}

\def\romlist{\begingroup\vskip10pt\romnum=1
\parskip=2pt\parindent=0pt \leftskip=5pc
\everypar{\llap{{\rm\romenumerate\hskip1em}}\advance\romnum by1}}



\long\def\fig#1#2#3{\vbox{\vskip1pc\vskip#1
\prevdepth=12pt \baselineskip=12pt
\vskip1pc
\hbox to\hsize{\hfill\vtop{\hsize=25pc\noindent{\eightbf Figure #2\ }
{\eightpoint#3}}\hfill}}}

\long\def\widefig#1#2#3{\vbox{\vskip1pc\vskip#1
\prevdepth=12pt \baselineskip=12pt
\vskip1pc
\hbox to\hsize{\hfill\vtop{\hsize=28pc\noindent{\eightbf Figure #2\ }
{\eightpoint#3}}\hfill}}}

\long\def\table#1#2{\vbox{\vskip0.5pc
\prevdepth=12pt \baselineskip=12pt
\hbox to\hsize{\hfill\vtop{\hsize=25pc\noindent{\eightbf Table #1\ }
{\eightpoint#2}}\hfill}}}

 
\def\rightheadline#1{\headline{\tenrm\hfil #1}}


\long\def\leftfig#1#2{\vbox{\smskip\hsize=220pt
\vtop{{\noindent {\bf #1}}}
\smskip
\noindent
\vbox{{\noindent #2}}
}}

\long\def\rightfig#1#2#3{\vbox{\smskip\vskip#1
\prevdepth=12pt \baselineskip=12pt
\hsize=210pt
\smskip
\vbox{\noindent{\eightbold #2}
\hskip1em{\eightpoint#3}}
}}

\long\def\concept#1#2#3#4#5{\bigskip\hrule
\vbox{\hbox{\leftfig{#1}{#2} \hskip3em
\rightfig{#3}{#4}{#5}} \smskip}
\hrule\bigskip}


\long\def\bconcept#1#2#3#4#5#6#7{
\vbox{
\hbox to \hsize{\vtop{\par #1}}
\concept{#2}{#3}{#4}{#5}{#6}
\hbox to \hsize{\vtop{\par #7}}
\smskip}
}




\def\boxit#1{\vbox{\hrule\hbox{\vrule\kern3pt
                                \vbox{\kern3pt#1\kern3pt}\kern3pt\vrule}\hrule}}
\def\centerboxit#1{$$\vbox{\hrule\hbox{\vrule\kern3pt
                                \vbox{\kern3pt#1\kern3pt}\kern3pt\vrule}\hrule}$$}

\long\def\boxtext#1#2{$$\boxit{\vbox{\hsize #1\noindent\strut #2\strut}}$$}

%
%
%

\def\picture #1 by #2 (#3){
  \vbox to #2{
    \hrule width #1 height 0pt depth 0pt
    \vfill
    \special{picture #3} 
    }
  }

\def\scaledpicture #1 by #2 (#3 scaled #4){{
  \dimen0=#1 \dimen1=#2
  \divide\dimen0 by 1000 \multiply\dimen0 by #4
  \divide\dimen1 by 1000 \multiply\dimen1 by #4
  \picture \dimen0 by \dimen1 (#3 scaled #4)}
  }

%
%

\long\def\captfig#1#2#3#4#5{\vbox{\vskip1pc
\hbox to\hsize{\hfill{\picture #1 by #2 (#3)}\hfill}
\prevdepth=9pt \baselineskip=9pt
\vskip1pc
\hbox to\hsize{\hfill\vtop{\hsize=24pc\noindent{\eightbold Figure #4}
\hskip1em{\eightpoint#5}}\hfill}}}

%
%
%

\def\illustration #1 by #2 (#3){
  \vskip#2\hskip#1\special{illustration #3} 
    }

\def\scaledillustration #1 by #2 (#3 scaled #4){{
  \dimen0=#1 \dimen1=#2
  \divide\dimen0 by 1000 \multiply\dimen0 by #4
  \divide\dimen1 by 1000 \multiply\dimen1 by #4
  \illustration \dimen0 by \dimen1 (#3 scaled #4)}
  }


\newbox\graybox
\newdimen\xgrayspace
\newdimen\ygrayspace
%
%
%
%
%
%
%
%
%

\def\Textshade#1#2#3#4#5#6{%
    \xgrayspace=#4pt%
    \ygrayspace=#4pt%
    \def\grayshade{#3}%
    \def\linewidth{#5}%
    \def\theradius{#6}%
    \setbox\graybox=\hbox{\surroundboxa{#2}}%
    \hbox{%
    \hbox to 0pt{%
    \PScommands
}%
    \box\graybox}}%
%
%
\long%

\long%
\def\Parashade#1#2#3#4#5#6#7{%
    \xgrayspace=#4pt%
    \ygrayspace=#4pt%
    \def\grayshade{#3}%
    \def\linewidth{#5}%
    \def\theradius{#6}%
    \def\thevskip{#7pt}%
    \setbox\graybox=\hbox{\surroundboxb{#2}}%
    \vskip\thevskip%
    \hbox{%
    \hbox to 0pt{%
    \PScommands
}%
     \box\graybox}%
     \vskip\thevskip%
}%
%
%
%
\long\def\surroundboxa#1{\leavevmode\hbox{\vtop{%
\vbox{\kern\ygrayspace%
\hbox{\kern\xgrayspace#1%
      \kern\xgrayspace}}\kern\ygrayspace}}}
%
%
\long\def\surroundboxb#1{\leavevmode\hbox{\vtop{%
\vbox{\kern\ygrayspace%
\hbox{\kern\xgrayspace\vbox{\advance\hsize-2\xgrayspace#1}%
      \kern\xgrayspace}}\kern\ygrayspace}}}
%
%
%
\long\def\PScommands{%
\special{rawpostscript
/sharpbox{%
           newpath
           xmin ymin moveto
           xmin ymax lineto
           xmax ymax lineto
           xmax ymin lineto
           xmin ymin lineto
           closepath 
          }bind def
}%
\special{rawpostscript
/sharpboxnb{%
           newpath
           xmin ymin moveto
           xmin ymax lineto
           xmax ymax lineto
           xmax ymin lineto
          }bind def
}%
\special{rawpostscript
/sharpboxnt{%
           newpath
           xmin ymax moveto
           xmin ymin lineto
           xmax ymin lineto
           xmax ymax lineto
          }bind def
}%
\special{rawpostscript
/roundbox{%
           newpath
           xmin radius add ymin moveto
           xmax ymin xmax ymax radius arcto
           xmax ymax xmin ymax radius arcto
           xmin ymax xmin ymin radius arcto
           xmin ymin xmax ymin radius arcto 16 {pop} repeat
           closepath
          }bind def
}%
\special{rawpostscript
/sharpcorners{%
               sharpbox gsave grayshade setgray fill grestore 
               linewidth setlinewidth stroke
              }bind def
}%
\special{rawpostscript
/sharpcornersnt{%
               sharpboxnt gsave grayshade setgray fill grestore 
               linewidth setlinewidth stroke
              }bind def
}%
\special{rawpostscript
/sharpcornersnb{%
               sharpboxnb gsave grayshade setgray fill grestore 
               linewidth setlinewidth stroke
              }bind def
}%
\special{rawpostscript
/roundcorners{%
               roundbox gsave grayshade setgray fill grestore 
               linewidth setlinewidth stroke
              }bind def
}%
\special{rawpostscript
/plainbox{%
           sharpbox grayshade setgray fill 
          }bind def
}%
%
\special{rawpostscript
/roundnoframe{%
               roundbox grayshade setgray fill 
              }bind def
}%
\special{rawpostscript
/sharpnoframe{%
               sharpbox grayshade setgray fill 
              }bind def
}%
}%
%
%

\def\pshade#1{\Parashade{sharpcorners}{#1}{0.95}{10}{0.5}{10}{10}}


\def\boxit#1{\vbox{\hrule\hbox{\vrule\kern3pt
                                \vbox{\kern3pt#1\kern3pt}\kern3pt\vrule}\hrule}}

\def\boxitnb#1{\vbox{\hrule\hbox{\vrule\kern3pt
                                \vbox{\kern3pt#1\kern3pt}\kern3pt\vrule}}}

\def\boxitnt#1{\vbox{\hbox{\vrule\kern3pt
                                \vbox{\kern3pt#1\kern3pt}\kern3pt\vrule}\hrule}}

\long\def\boxtext#1#2{$$\boxit{\vbox{\hsize #1\noindent\strut #2\strut}}$$}



\def\texshopbox#1{\boxtext{462pt}{\vskip-1.5pc\pshade{\vskip-1.0pc#1\vskip-2.0pc}}}


%
%
%
%
%
%
%
%
\font\helbigbig=cmr10 scaled 2500%
\font\helbigb=cmbx10 scaled 1500%
\font\eightbold=cmbx8%

\def\tenf{\hel}%
\def\tenit{\heli}%
\def\ninef{\ninehel}%
\def\nineit{\nineheli}%
%
%


\font\tenrm=cmr10%
\font\teni=cmmi10%
\font\tensy=cmsy10%
\font\tenbf=cmbx10%
\font\tentt=cmtt10%
\font\tenit=cmti10%
\font\tensl=cmsl10%

\def\tenpoint{\def\rm{\fam0\tenrm}%
\textfont0=\tenrm%
\textfont1=\teni%
\textfont2=\tensy%
\textfont\itfam=\tenit%
\textfont\slfam=\tensl%
\textfont\ttfam=\tentt%
\textfont\bffam=\tenbf%
\scriptfont0=\sevenrm%
\scriptfont1=\seveni%
\scriptfont2=\sevensy%
\scriptscriptfont0=\sixrm%
\scriptscriptfont1=\sixi%
\scriptscriptfont2=\sixsy%
\def\it{\fam\itfam\tenit}%
\def\tt{\fam\ttfam\tentt}%
\def\sl{\fam\slfam\tensl}%
\scriptfont\bffam=\sevenbf%
\scriptscriptfont\bffam=\sixbf%
\def\bf{\fam\bffam\tenbf}%
\normalbaselineskip=18pt%
\normalbaselines\rm}%

\font\ninerm=cmr9%
\font\ninebf=cmbx9%
\font\nineit=cmti9%
\font\ninesy=cmsy9%
\font\ninei=cmmi9%
\font\ninett=cmtt9%
\font\ninesl=cmsl9%

\def\ninepoint{\def\rm{\fam0\ninerm}%
\textfont0=\ninerm%
\textfont1=\ninei%
\textfont2=\ninesy%
\textfont\itfam=\nineit%
\textfont\slfam=\ninesl%
\textfont\ttfam=\ninett%
\textfont\bffam=\ninebf%
\scriptfont0=\sixrm%
\scriptfont1=\sixi%
\scriptfont2=\sixsy%
\def\it{\fam\itfam\nineit}%
\def\tt{\fam\ttfam\ninett}%
\def\sl{\fam\slfam\ninesl}%
\scriptfont\bffam=\sixbf%
\scriptscriptfont\bffam=\fivebf%
\def\bf{\fam\bffam\ninebf}%
\normalbaselineskip=16pt%
\normalbaselines\rm}%

\font\eightrm=cmr8%
\font\eighti=cmmi8%
\font\eightsy=cmsy8%
\font\eightbf=cmbx8%
\font\eighttt=cmtt8%
\font\eightit=cmti8%
\font\eightsl=cmsl8%

\def\eightpoint{\def\rm{\fam0\eightrm}%
\textfont0=\eightrm%
\textfont1=\eighti%
\textfont2=\eightsy%
\textfont\itfam=\eightit%
\textfont\slfam=\eightsl%
\textfont\ttfam=\eighttt%
\textfont\bffam=\eightbf%
\scriptfont0=\sixrm%
\scriptfont1=\sixi%
\scriptfont2=\sixsy%
\scriptscriptfont0=\fiverm%
\scriptscriptfont1=\fivei%
\scriptscriptfont2=\fivesy%
\def\it{\fam\itfam\eightit}%
\def\tt{\fam\ttfam\eighttt}%
\def\sl{\fam\slfam\eightsl}%
\scriptscriptfont\bffam=\fivebf%
\def\bf{\fam\bffam\eightbf}%
\normalbaselineskip=14pt%
\normalbaselines\rm}%

\font\sevenrm=cmr7%
\font\seveni=cmmi7%
\font\sevensy=cmsy7%
\font\sevenbf=cmbx7%

\font\sixrm=cmr6%
\font\sixi=cmmi6%
\font\sixsy=cmsy6%
\font\sixbf=cmbx6%

\fontdimen13\tensy=2.6pt%
\fontdimen14\tensy=2.6pt%
\fontdimen15\tensy=2.6pt%
\fontdimen16\tensy=1.2pt%
\fontdimen17\tensy=1.2pt%
\fontdimen18\tensy=1.2pt%

\def\tenf{\tenpoint}%
\def\ninef{\ninepoint}%
%




\def\texshopbox#1{\boxtext{462pt}{\vskip-1.5pc\pshade{\vskip-1.0pc#1\vskip-2.0pc}}}


\input miniltx

\ifx\pdfoutput\undefined
  \def\Gin@driver{dvips.def} 
\else
  \def\Gin@driver{pdftex.def} 
\fi

\input graphicx.sty
\resetatcatcode

\long\def\fig#1#2#3{\vbox{\vskip1pc\vskip#1
\prevdepth=12pt \baselineskip=12pt
\vskip1pc
\hbox to\hsize{\hskip3pc\hfill\vtop{\hsize=35pc\noindent{\eightbf Figure #2\ }
{\eightpoint#3}}\hfill}}}

\def\show#1{}

\def\frac#1#2{{#1\over #2}}

\rightheadline{\botmark}

\pageno=1

\rightheadline{\botmark}

\pn {\bf October 2011 (Revised Feb.\ 2012)}\hfill {\bf Report LIDS - 2874}
\bigskip 
\bigskip\bigskip

\bigskip\bigskip

\def\longpapertitle#1#2#3{{\bf \centerline{\helbigb
{#1}}}\medskip{\bf \centerline{\helbigb
{#2}}}\medskip{\centerline{
by}}\medskip{\bf \centerline{
{#3}}}\bigskip}

\def\longpapertitle#1#2#3{{\bf \centerline{\helbigb
{#1}}}\medskip{\bf \centerline{\helbigb
{#2}}}\medskip{\centerline{
}}\medskip{\bf \centerline{
{#3}}}\bigskip}

\vskip-3pc
\longpapertitle{Lambda-Policy Iteration: A Review and a}{New Implementation\footnote{\dag}
{\ninepoint  To appear in {\it Reinforcement Learning and Approximate Dynamic Programming for Feedback Control\/}, by F.\ Lewis and D.\ Liu (eds.), IEEE Press Computational Intelligence Series.}
}
{Dimitri P.\ Bertsekas
\footnote{\ddag}
{\ninepoint  The author is with the Dept.\ of Electr.\ Engineering and
Comp.\ Science, M.I.T., Cambridge, Mass., 02139. His research was supported by NSF Grant ECCS-0801549, and by the Air Force Grant FA9550-10-1-0412. Thanks are due to Bruno Scherrer for helpful comments, to Huizhen Yu for related collaboration, and to Mengdi Wang for assistance with computational experimentation.}
}

\centerline{\bf Abstract}
In this paper we discuss $\l$-policy iteration, a method for exact and approximate dynamic programming. It is intermediate between the classical value iteration (VI) and policy iteration (PI) methods, and it is closely related to optimistic (also known as modified) PI, whereby each policy evaluation is done approximately, using a finite number of VI. We review the theory of the method and associated questions of bias and exploration arising in simulation-based cost function approximation. We then discuss various implementations, which offer advantages over well-established PI methods that use LSPE($\l$), LSTD($\l$), or TD($\l$) for policy evaluation with cost function approximation. One of these implementations is based on a new simulation scheme, called geometric sampling, which uses multiple short trajectories rather than a single infinitely long trajectory. 


\def\tl{\tilde}
\def\ol{\bar}

\def\old#1{}

\vskip  -4mm
\section{Introduction}
\vskip  -2mm

\pn Approximate dynamic programming  (DP for short) has attracted substantial research interest, and has a wide range of applications, because of its potential to address large and complex problems that may not be treatable in other ways.  The literature on the subject is very extensive, and includes several textbooks, research monographs, and surveys that relate to the computational context of this paper. For a nonexhaustive list, we mention the books by Bertsekas and Tsitsiklis [BeT96], Sutton and Barto [SuB98], Gosavi [Gos03], Cao [Cao07], Chang, Fu, Hu, and Marcus [CFH07], Meyn [Mey07], Powell [Pow07], Borkar [Bor08], Haykin [Hay08], Busoniu, Babuska, De Schutter, and Ernst [BBD10], and the author's text in preparation [Ber11a]; the edited volumes and special issues by White and Sofge [WhS92], Si, Barto, Powell, and Wunsch [SBP04], Lewis, Lendaris, and Liu [LLL08], and the 2007-2009 Proceedings of the IEEE Symposium on Approximate Dynamic Programming and Reinforcement Learning; and the recent surveys by Borkar [Bor09], Lewis and Vrabie [LeV09], Werbos [Wer09], Szepesvari [Sze10], and Bertsekas [Ber11b].

The purpose of this paper is to critically review and extend a class of methods for exact and approximate DP, which are based on the $\l$-policy iteration ($\l$-PI) method, proposed by Bertsekas and Ioffe [BeI96]. This method is intermediate between the classical value iteration (VI) and policy iteration (PI) methods, and it is closely related to optimistic (also known as modified) PI, whereby each policy evaluation is done approximately, using a finite number of VI. It was originally used as the starting point for the development of approximate simulation-based DP methods of the temporal difference (TD) type, such as LSPE($\l$) (see [BeI96], and also [BeT96], Sections 2.3.1 and 8.3). The emphasis in this paper is on implementations of $\l$-PI, which provide alternatives to approximate PI methods that use other more established methods for policy evaluation.

We will focus on the $\a$-discounted $n$-state Markovian Decision Problem (MDP), although the main ideas are more broadly applicable. The problem involves states $1,\ldots,n$, controls $u\in U(i)$ at state $i$, transition probabilities $p_{ij}(u)$, and cost $g(i,u,j)$ for transition from $i$ to $j$ under control $u$. A (stationary) policy $\m$ is a function from states $i$ to admissible controls $u\in U(i)$, and $J_\m(i)$ is the cost starting from state $i$ and using policy $\m$. It is well-known (see e.g., Puterman [Put94] or Bertsekas [Ber07]) that the vector $J_\m\in\rn$, which has components $J_\m(i)$,\footnote{\dag}{\ninepoint  In our notation, $\re^n$ is the $n$-dimensional Euclidean space, all vectors in $\re^n$ are viewed as column vectors, and a prime denotes transposition. The identity matrix is denoted by $I$.} is the unique fixed point of the mapping $T_\m:\rn\mapsto\rn$, which maps $J\in\rn$ to the vector $T_\m J\in\rn$ that has components 
$$(T_\m J)(i)=\sum_{j=1}^np_{ij}\big(\m(i)\big)\big(g(i,\m(i),j)+\a J(j)\big),\qquad i=1,\ldots,n.\eqnum\show{oneo}$$
Similarly, the optimal costs starting from $i=1,\ldots,n$, are denoted  $J^*(i)$, and the optimal cost vector $J^*\in\rn$, which has components $J^*(i)$,  is the unique fixed point of the mapping $T:\rn\mapsto\rn$ defined by
$$(T J)(i)=\min_{u\in U(i)}\sum_{j=1}^np_{ij}(u)\big(g(i,u,j)+\a J(j)\big),\qquad i=1,\ldots,n.\xdef\discountstar{\lab}\eqnum\show{oneo}$$
An important property is that $T_\m$ and $T$ are sup-norm contractions. In particular, the iterations  $J_{k+1}=T_\m J_k$ and $J_{k+1}=TJ_k$ converge to $J_\m$ and $J^*$, respectively, from any starting point $J_0$ - this is the VI method.

A major alternative to VI is PI. It  produces a sequence of policies and associated cost functions through iterations that have two phases: {\it policy evaluation} (where the cost function of a policy is evaluated), and {\it policy improvement}  (where a new policy is generated). In the exact form of the algorithm, the current policy  $\m$ is improved by finding $\ol \m$ that satisfies $T_{\ol \m}J_\m=TJ_\m$ [i.e., by minimizing in the right-hand side of Eq.\ \discountstar\ with $J_\m$ in place of $J$]. The improved policy $\ol\m$ is evaluated by solving the linear system of equations $J_{\ol\m}=T_{\ol\m}J_{\ol\m}$, and $(J_{\ol\m},\ol\m)$ becomes the new cost vector-policy pair, which is used to start a new iteration. Thus, the exact form of PI can be succinctly defined as
$$T_{\m_{k+1}}J_{k}=TJ_{k},\qquad J_{k+1}=T_{\m_{k+1}}J_{k+1},\eqnum\show{oneo}$$
with the equation on the left describing the policy improvement and the equation on the right describing the evaluation of ${\m_{k+1}}$.

In a variant of the method, a policy $\m_{k+1}$ is evaluated by a finite number of applications of $T_{\m_{k+1}}$ to an approximate evaluation of the preceding policy. This is known as ``optimistic" or ``modified" PI, and its motivation is that in problems with a large number of states, the linear system $J_{k+1}=T_{\m_{k+1}}J_{k+1}$ cannot be practically solved directly by matrix inversion, so it is best solved iteratively by VI. The method can be succinctly defined as
$$T_{{\m_{k+1}}}J_{k}=TJ_{k},\qquad J_{k+1}=T_{\m_{k+1}}^{m_k} J_{k}.\xdef\opteval{\lab}\eqnum\show{lsp}$$
If the number $m_k$ of applications of $T_{{\m_{k+1}}}$ is very large, the exact form of PI is essentially obtained, but practice has shown that it is most efficient to use a moderate value of $m_k$. In this case, the algorithm looks like a hybrid of VI and PI, involving a sequence of alternate applications of $T$ and $T_{\m_k}$, with $\m_k$ changing over time. Optimistic PI is generally believed to be more computationally efficient that either VI or PI. This is particularly so for problems where $n$ is very large and implementation of exact PI is difficult due to the associated $n\times n$ matrix inversion, and also for problems with a large number of controls, where the overhead due to minimization over all controls $u\in U(i)$ in the mapping $T$ [cf.\ Eq.\ \discountstar] is substantial. 

We note that the convergence properties of the optimistic PI method \opteval\ are quite complicated and have been the subject of continuing research. The convergence $J_k\to J^*$ has been established by Rothblum [Rot79] (see also the more recent work by Canbolat and Rothblum [CaR11], which extends some of the results of [Rot79]). On the other hand, when optimistic PI is implemented asynchronously (as it normally would be when simulation is used), it may oscillate as shown by the convergence counterexamples of Williams and Baird [WiB93]. Recent work of Bertsekas and Yu [BeY10a], [BeY10b], [YuB11] has developed convergent variants of synchronous and asynchronous optimistic PI and Q-learning, based on a new way to perform policy evaluation: by solving approximately an optimal stopping problem rather than a system of linear equations. 

The  $\l$-PI method is a form of optimistic PI, given by
$$T_{\m_{k+1}}J_{k}=TJ_{k},\qquad J_{k+1}=T_{\m_{k+1}}^{(\l)}J_{k},\xdef\lamdaeval{\lab}\eqnum\show{lsp}$$
where for any $\m$ and  $\l\in[0,1)$, $T_\m^{(\l)}$ is the linear mapping given by
$$T_\m^{(\l)}=(1-\l)\sum_{\ell=0}^\infty \l^\ell T_{\m}^{\ell+1}.\xdef\tlambdamap{\lab}\eqnum\show{oneo}$$
Note that the mapping $T_\m^{(\l)}$ is central in much recent research on approximate DP, simulation-based PI, and TD methods, as will be discussed in the sequel. 

To compare the optimistic PI method \opteval\ and the $\l$-PI method \lamdaeval, note that both mappings $T_{\m_{k+1}}^{m_k}$ and $T_{\m_{k+1}}^{(\l)}$ appearing in Eqs.\ \opteval\ and \lamdaeval, involve multiple applications of the VI mapping $T_{\m_{k+1}}$: a fixed number $m_k$ in the former case (with $m_k=1$ corresponding to VI and $m_k\to\infty$ corresponding to PI), and a geometrically weighted number  in the latter case (with $\l=0$ corresponding to VI and $\l\to1$ corresponding to PI). 
Thus optimistic PI and $\l$-PI are similar: they just control the accuracy of  the approximation $J_{k+1}\approx J_{\m_{k+1}}$ by applying VI in different ways. In a classical DP/non-simulation-based setting, $\l$-PI is far more complicated relative to optimistic PI, since exact computations using the mapping $T_{\m}^{(\l)}$ are unwieldy. However, this advantage of optimistic PI is dissipated in a simulation context, where computations involving $T_{\m}^{(\l)}$ can be performed conveniently, as extensive analytical and experimental work with TD methods has demonstrated.

Recent research on DP has focused on the use of simulation, in order to deal with model-free situations where the transition probabilities and/or the cost per stage are not known explicitly, and also to deal with the associated high-dimensional linear algebra operations. 
For problems with very large number of states, the evaluation of various fixed points of mappings, such as $T_\m$ or $T_\m^{(\l)}$, is typically done by approximation
 with a vector $\Phi r$ from the subspace $S=\{\Phi r\mid r\in\re^s\}$ that is spanned by the columns of an $n\times s$ matrix $\Phi$. In this paper we will focus on the  {\it projected equation approach\/}, whereby given a generic mapping $L:\rn\mapsto\rn$ (such as for example $T_\m$) we approximate its fixed point by solving the equation
 $$\Phi r=\Pi L(\Phi r),$$
where $\Pi$ denotes projection onto the subspace $S$.
The projection is with respect to a Euclidean norm $\|\cdot\|_\xi$, weighted by a suitable vector  $\xi$ of positive weights. An alternative possibility is to solve instead the equation
  $$\Phi r=\Pi L^{(\nu)}(\Phi r),\xdef\projeqonelambda{\lab}\eqnum\show{oneo}$$
  where, similar to Eq.\ \tlambdamap,
$$L^{(\nu)}=(1-\nu)\sum_{\ell=0}^\infty \nu^\ell L^{\ell+1},$$
and $\nu\in[0,1)$ is a parameter [not necessarily the same as the $\l$ parameter in Eqs.\ \lamdaeval-\tlambdamap].
In our context we will encounter several different types of mappings $L$, and in all cases $L$ is a contraction with respect to the projection norm $\|\cdot\|_\xi$, with fixed point $\hat J$, while $\Pi L^{(\nu)}$ are contractions with respect to $\|\cdot\|_\xi$ for all $\nu\in[0,1)$. It is well-known that the fixed point of $\Pi L^{(\nu)}$, denoted $\Phi r(\nu)$, converges to $\Pi \hat J$ as $\nu\to1$. The norm of the difference 
$\Phi r(\nu)-\Pi \hat J$
 is known as the {\it bias}. Its size/norm depends on $\nu$ and is generally smaller as $\nu$ gets closer to 1 (see [BeT96], [TsV97], [YuB10] for error bound analyses).

A common example of fixed point approximation in PI is when $L=T_\m$ for a policy $\m$, in which case the fixed point of $\Pi L$ or $\Pi L^{(\nu)}$ is an approximation to the fixed point of $T_{\m}$, i.e., the cost vector $J_\m$. If the Markov chain corresponding to $\m$ is irreducible and $\xi$ is the corresponding steady-state distribution vector, the mapping $\Pi T^{(\l)}_\m$ is a contraction with respect to $\|\cdot\|_\xi$ for all $\l\in[0,1)$, and is unique fixed point, denoted $\Phi r_\m(\l)$, converges to $\Pi J_\m$ as $\l\to1$.
Generally, the projected equation $\Phi r=\Pi T^{(\l)}_\m(\Phi r)$ is solved by a simulation process that generates a sequence of states according to a sampling scheme to be discussed later, and then by matrix inversion [this is the Least Squares Temporal Differences [LSTD($\l$)] method, proposed by Bradtke and Barto [BrB96]], or by iteration, using the TD($\l$) method, proposed by Sutton [Sut87] and analyzed by Tsitsiklis and VanRoy [TsV97] among others, or the Least Squares Policy Evaluation [LSPE($\l$)] method, proposed by Bertsekas and Ioffe [BeI96].\footnote{\dag}{\ninepoint  The paper [BeI96] as well as the book [BeT96] used the name ``$\l$-policy iteration" for both the lookup table and the compact representation versions of the method described here, and tested a compact representation version on the game of tetris, a challenging SSP problem. The name ``LSPE" was first used in the subsequent paper by Nedi\'c and Bertsekas [NeB03] to describe a specific  iterative implementation of the $\l$-PI method with cost function approximation for discounted MDP (essentially the discounted version of the implementation used in [BeI96] and [BeT96] for the aforementioned tetris case study). Reference [NeB03] proved convergence of the LSPE($\l$) method, as described in Section 3.1, for the case of a diminishing stepsize. Convergence for a stepsize equal to 1 was proved shortly afterwards by Bertsekas,  Borkar,  and Nedi\'c [BBN04].  The use of two different names for essentially the same method has been a source of some confusion. While in practical implementations these two names refer to algorithms that are closely related, we reserve the name ``$\l$-policy iteration" for the more abstract form \lamdaeval-\tlambdamap, and we will view LSPE($\l$) as an implementation of $\l$-PI (see Section 4.1).} These methods are extensively discussed in the literature, and exhibit complex and sometimes pathological behavior, particularly when embedded within PI (see [Ber95], [SzL06], [ThS09] for some notable failures, and [Ber10] for a recent assessment). Moreover matrix inversion and iterative methods, like TD($\l$), LSTD($\l$), and LSPE($\l$), can be used for solving  not only the projected equation $\Phi r=\Pi T^{(\l)}_\m(\Phi r)$, but also the more general equation $\Phi r=\Pi L^{(\nu)}(\Phi r)$ of Eq.\ \projeqonelambda, as long as $L$ is a linear mapping that is convenient for the use of simulation [and in the case of TD($\l$) and LSPE($\l$), $\Pi L^{(\nu)}$ is a contraction; see [BeY09] or [Ber11c]].
  
In this paper we will review some of the basic issues in approximate PI using the projected equation approach, thereby setting the stage for assessing the relative strengths and weaknesses of the $\l$-PI methodology. We will then focus on three alternative implementations of $\l$-PI, which involve simulation and cost function approximation. The first  is basically the LSPE($\l$) method as implemented in [BeI96]. The second is an interesting recent proposal by Thiery and Scherrer [ThS10a], who gave extensive and quite successful computational results, as well as error bounds [ThS10b]. The third implementation is new and may have some advantages over the first two. We will argue that it deals better with the combined issues of bias and exploration. This implementation embodies a new idea for $\l$-methods: a simulation scheme, called {\it geometric sampling\/},  that uses multiple short trajectories with random geometrically distributed length, and exploration-enhanced restart, rather than a single infinitely long trajectory. 

The three implementations are described in Section 4, following a discussion of the generic properties of exact $\l$-PI in Section 2, and the LSTD($\l$) and LSPE($\l$) methods in Section 3. In our description, these implementations are model-based and use cost function approximation, but there are versions that are model-free and use Q-factor approximation; these can be straightforwardly constructed by the reader.

\section{Lambda-Policy Iteration without Cost Function Approximation}

\pn We first recall a central result  from [BeI96]. It provides a helpful characterization of the $\l$-PI method \lamdaeval,  which will later become the basis for cost function approximations.

\xdef\proplpolt{\propn}\propnum\show{myproposition}

\texshopbox{
\proposition{\proplpolt:} Given $\l\in[0,1)$, $J_k$, and $\m_{k+1}$, consider the
mapping $W_k$ defined by
$$W_kJ=(1-\l)T_{\m_{k+1}}J_k+\l
T_{\m_{k+1}}J.\xdef\mapm{\lab}\eqnum\show{mapm}$$
\nitem{(a)} $W_k$ is a sup-norm contraction of modulus $\l\a$.\nitem{(b)} The vector $J_{k+1}=T_{\m_{k+1}}^{(\l)}J_{k}$ generated next by the $\l$-PI method \lamdaeval\
is the unique fixed point of $W_k$.}

\proof (a) For any two vectors $J$ and $\ol J$, using the definition \mapm\ of $W_k$, we
have
$$\|W_kJ-W_k\ol J\|= \bl\|\l(T_{\m_{k+1}}J-T_{\m_{k+1}}\ol J)\br\|= \l\|T_{\m_{k+1}}J-T_{\m_{k+1}}\ol J\|\le \l\a\|J-\ol J\|,$$
where  $\|\cdot\|$ denotes the sup-norm, so $W_k$ is a sup-norm contraction with modulus $\l\a$. 
\smskip
\pn (b) We have
$$J_{k+1}=T_{\m_{k+1}}^{(\l)}J_{k}=(1-\l)\sum_{\ell=0}^\infty \l^\ell T_{\m_{k+1}}^{\ell+1}J_k,$$
so the fixed point property to be shown, $J_{k+1}=W_k J_{k+1}$, is written as
$$(1-\l)\sum_{\ell=0}^\infty \l^\ell T_{\m_{k+1}}^{\ell+1}J_k= (1-\l)T_{\m_{k+1}}J_k+\l
T_{\m_{k+1}}(1-\l)\sum_{\ell=0}^\infty \l^\ell T_{\m_{k+1}}^{\ell+1}J_k,$$
and evidently holds. \qed

From part (b) of the preceding proposition, we see that $J_{k+1}=W_kJ_{k+1}$, or equivalently
$$J_{k+1}(i)=\sum_{j=1}^np_{ij}\big(\m_{k+1}(i)\big)\Big(g\big(i,\m_{k+1}(i),j\big)+(1-\l)\a J_{k}(j)+\l\a J_{k+1}(j)\Big),\qquad i=1,\ldots,n.\xdef\beleqstop{\lab}\eqnum\show{mkm}$$
The solution of this fixed point equation can be obtained by viewing it as Bellman's equation for two equivalent MDP. 
\nitem{(a)} {\it As Bellman's equation for an infinite-horizon $\l\a$-discounted MDP\/} where $\m_{k+1}$ is the only policy, and the cost per stage is
$$g \big(i,\m_{k+1}(i),j\big)+(1-\l)\a J_{k}(j).$$
\nitem{(b)} 
{\it As Bellman's equation for an infinite-horizon stopping problem\/} where $\m_{k+1}$ is the only policy. In particular, $J_{k+1}$ is the cost vector of policy $\m_{k+1}$ in a stopping problem that is derived from the given $\a$-discounted problem by introducing transitions from each state $j$ to an artificial termination state as follows: at state $i$ we first make a transition to $j$ with probability $p_{ij}\big(\m_{k+1}(i)\big)$ and transition cost $g\big(i,\m_{k+1}(i),j\big)$; then we either stay at $j$ and wait for the next transition (this occurs with probability $\l$), or else we move from $j$ to the termination state with an additional termination cost $\a J_k(j)$  (this occurs with probability $1-\l$). All transition costs as well as the termination cost are discounted by an additional factor $\a$ with each transition. 
\smskip 

The convergence and rate of convergence of the $\l$-PI method \lamdaeval\ was given in [BeI96] and also in [BeT96], Prop.\ 2.8. We will simply quote the results for completeness.

\xdef\proplpolth{\propn}\propnum\show{myproposition}

\texshopbox{\proposition{\proplpolth:}  Assume that 
 $\l\in[0,1)$, and let $\{J_k,\m_k\}$ be the sequence
generated by the $\l$-PI method \lamdaeval. 
Then  $J_k$ converges to
$J^*$. Furthermore, for all $k$ greater than some index $\ol k$, $\m_k$ is optimal.}

\xdef\proplpoltha{\propn}\propnum\show{myproposition}

\texshopbox{\proposition{\proplpoltha:} Let the assumptions of Prop.\ \proplpolth\ hold and let  $\ol k$ be the index such that for all $k\ge \ol k$, $\m_k$ is optimal. The sequence $\{J_k\}$ 
generated by the $\l$-PI method \lamdaeval\ satisfies for all $k>\ol k$
$$\|J_{k+1}-J^*\|\le {\a(1-\l)\over
1-\l\a}\|J_k-J^*\|,\xdef\conrate{\lab}\eqnum\show{conr}$$
where $\|\cdot\|$ denotes the sup-norm.}

Note that the convergence rate estimate \conrate\ holds only for $k\ge \ol k$, essentially after an optimal policy has been identified, as per Prop.\ \proplpolth. Nonetheless, this rate estimate is qualitatively correct, and supports the empirical observation that the iterates $(J_k,\m_k)$ generated by $\l$-PI converge faster as $\l$ increases. Indeed in the limit, as $\l\to1$, $\l$-PI becomes exact PI, and converges to the optimum in a finite number of iterations. On the other hand, the computation of $J_{k+1}=T_{\m_{k+1}}^{(\l)}J_{k}$ [cf.\ Eq.\ \lamdaeval] becomes more time-consuming as $\l$ increases, particularly when simulation is used, because the simulation-based calculation of $T_{\m_{k+1}}^{(\l)}J_{k}$ involves more simulation noise as $\l$ gets larger.

We finally note that Props.\ \proplpolth\ and \proplpoltha\ apply to synchronous implementations of $\l$-PI.  When implemented asynchronously, $\l$-PI has similar convergence difficulties to optimistic PI. To see this, note that asynchronous implementations of these two methods essentially coincide when $m_k=1$ in Eq.\ \lamdaeval\ and $\l=0$ in Eq.\ \opteval, and the counterexamples of Williams and Baird [WiB93] apply. Thus the development of convergent versions of asynchronous $\l$-PI is an open research question.

\vskip-1.5pc

\section{Approximate Policy Evaluation Using Projected Equations}

\vskip-0.5pc

\pn In PI methods with cost function approximation, we evaluate $\m$ by approximating $J_\m$ with a vector $\Phi r_\m$ from the subspace $S=\{\Phi r\mid r\in\re^s\}$, spanned by the columns of an $n\times s$ matrix $\Phi$, which may be viewed as basis functions.
We  generate an ``improved" policy $\ol \m$ using the formula $T_{\ol \m}(\Phi r_\m)=T(\Phi r_\m)$, i.e., 
$$\ol \m(i)\in\arg\min_{u\in U(i)}\sum_{j=1}^n p_{ij}(u)\bl(g(i,u,j)+\a\phi(j)'r_\m\br),\qquad i=1,\ldots,n,$$
where $\phi(j)'$ is the row of $\Phi$ that corresponds to state $j$ [the method terminates with $\m$ if $T_{\m}(\Phi r_{\m})=T(\Phi r_{\m})$]. We then repeat with $\m$ replaced by $\ol \m$. For the purposes of this paper, we assume that $\Phi$ has rank $s$, and that the Markov chain corresponding to $\m$ is irreducible. 

As noted earlier, in the projected equation  approach to approximate PI, we approximate $J_\m$ with a vector of the form $\Phi r_\m(\l)$ that solves the fixed point problem
 $$\Phi r=\Pi T_\m^{(\l)}(\Phi r).\xdef\projeqone{\lab}\eqnum\show{oneo}$$
Here $\Pi$ denotes projection onto the subspace $S$ with respect to a weighted Euclidean norm $\|\cdot\|_\xi$, where $\xi=(\xi_1,\ldots,\xi_n)$ is a probability distribution with positive components (i.e., $\|J\|_\xi^2=\sum_{i=1}^n\xi_ix_i^2$, where $\xi_i>0$ for all $i$). In nonoptimistic PI methods, the projected equation \projeqone\ is solved exactly, while in optimistic PI methods it is solved approximately. We note that this approach has a long history in the context of Galerkin methods for the approximate solution of high-dimensional or infinite-dimensional linear equations (partial differential, integral, inverse problems, etc; see  e.g., [Kra72], [Fle84]). In fact some of the policy evaluation theory referred to in this paper applies to general projected equations arising in contexts beyond DP (see [BeY09], [Ber09], [Yu10a,b], [Ber11c]). However, Monte Carlo simulation is not part of the Galerkin methodology, as currently practiced in the numerical analysis field. For this reason much of the extensive available knowledge about Galerkin methods does not apply to the approximate DP context, which is primarily simulation-oriented.

We now discuss some of the issues relating to projected equations. While we focus on Eq.\ \projeqone, much of our discussion also applies to the more general projected equations.

\subsubsection{Exploration-Contraction Tradeoff}

\pn An important choice in the projected equation approach is the distribution $\xi$ that defines the projection norm $\|\cdot\|_\xi$. This distribution is sometimes chosen to be the steady-state probability vector $\xi_\m$ of the Markov chain corresponding to $\m$, in which case the mapping $\Pi T_\m^{(\l)}$ can be shown to be a contraction with respect to $\|\cdot\|_{\xi_\m}$ with modulus
$$\a_\l={\a(1- \l)\over 1-\l\a}\xdef\alambda{\lab}\eqnum\show{lsp}$$
(see [BeT96], Lemma 6.6, or [Ber07], Prop.\ 6.3.3). 

On the other hand the choice of $\xi$ is related to  {\it exploration}, i.e., the need to collect an adequately rich set of samples from a broad and representative set of states. This is a critical issue in simulation-based PI, and results in a well-known tradeoff: to evaluate a policy
$\m$, we may need to generate cost samples using $\m$, but this may affect the simulation results
by underrepresenting states that are unlikely to occur under  $\m$ (more weight is placed on states that are visited more frequently under $\m$). As a result, the
cost-to-go estimates of the underrepresented states may be highly inaccurate, causing potentially serious
errors in the calculation of the improved control policy.

A well-known approach for exploration is to choose $\xi$ to be a mixture of the form
$$\xi=(1-\b)\xi_\m+\b \tl \xi,\xdef\explore{\lab}\eqnum\show{lsp}$$
where $\b\in(0,1)$ and $\tl\xi$ is another distribution (often referred to as the {\it off-policy} distribution), which is added to enhance exploration (see the discussion of Section 1). Unfortunately, with such a choice the contraction property of $\Pi T_\m^{(\l)}$ comes into doubt: it depends on the size of the parameters $\l$ and $\b$ [it can be shown that $\Pi T_\m^{(\l)}$ is a contraction for any $\b\in[0,1)$ provided $\l$ is close enough to 1, and it is a contraction for any $\l\in[0,1)$ provided $\b$ is close enough to 0]. This is important because for convergence of iterative methods such as TD($\l$) and some forms of LSPE($\l$), it is critical that $\Pi T_\m^{(\l)}$ be a contraction. Thus there is a tradeoff between exploration enhancement using the mixture distribution \explore\ and ability to use a broader range of methods for solution of the projected equation.

\subsubsection{Bias}

\pn  While the Bellman equation $J=T_\m^{(\l)}J$ has the same fixed point $J_\m$ for all $\l\in[0,1)$, the fixed point $\Phi r_\m(\l)$ of the projected version \projeqone\ depends on $\l$. The difference of $\Phi r_\m(\l)$ and the closest point of $S$ to $J_\m$, $\Phi r_\m(\l)-\Pi J_\m$, is generally nonzero.  Its norm,  the bias, tends to decrease to 0 as $\l\uparrow 1$ and tends to increase as $\l\downarrow 0$. It is known that the bias can be very large and may seriously degrade the practical value of the approximate policy evaluation for small values of $\l$; see [Ber95] for some examples.

The following is a well-known error bound for the case $\xi={\xi_\m}$:
$$\|J_\m-\Phi r_\m(\l)\|_{\xi_\m}\le {1\over \sqrt{1-\a_\l^2}}\,\|J_\m-\Pi
J_\m\|_{\xi_\m},\xdef\llspeest{\lab}\eqnum\show{lsp}$$
where $\a_\l$ is given by Eq.\ \alambda, and $\|\cdot\|_{\xi_\m}$ is the weighted Euclidean norm corresponding to
$\xi={\xi_\m}$, the steady-state probability vector of the Markov chain corresponding to $\m$.
Thus the error bound becomes worse as $\l$ decreases (and $\a_\l$ increases), suggesting a larger size of bias. While the bound is rather conservative, the paper by Yu and Bertsekas [YuB10] (see also Scherrer [Sch10]) derives sharper error bounds, which also apply to cases where $\xi\ne {\xi_\m}$ and $\Pi T_\m^{(\l)}$ is not a contraction. These error bounds and the bound \llspeest\  are consistent in suggesting that the bias increases as $\l$ decreases, and they are also largely consistent with the results of computational experimentation.

\subsubsection{Bias-Variance Tradeoff}

\pn In simulation-based methods for solving the projected equation \projeqone, one must deal with the effects of simulation error. Generally as $\l$ increases, the methods become more vulnerable to simulation noise, and hence require more sampling for good performance.  Indeed, the
noise in a simulation sample of an $\ell$-stages cost vector $T^\ell_\m J$ tends to be larger as $\ell$ increases, and from the
formula
$$T^{(\l)}_\m=(1-\l)\sum_{\ell=0}^\infty\l^\ell T^{\ell+1}_\m$$
it can be seen that simulation samples of 
$T^{(\l)}_\m(\Phi r_k)$ tend to contain more noise as $\l$ increases. This is consistent with practical experience, and gives rise to the so called bias-variance tradeoff: a large value of $\l$ to reduce bias results in slower and less reliable computation because of higher simulation noise (and consequently, a larger number of samples to achieve the same accuracy of various simulation-based estimates). Generally, there is no rule of thumb for selecting $\l$, which is usually chosen with
some trial and error.

In summary, the preceding discussion suggests that if simulation noise is not an issue (i.e., one can afford many simulation samples) one should choose large values of $\lambda$, since then the bias is reduced and one may afford greater exploration without losing the contraction property of $\Pi T_\m^{(\l)}$. In the contrary case, however, the degradation of the estimate of $J_\m$ due to simulation noise may offset whatever bias/contraction benefits a large value of $\l$ may bring.

\subsection{TD Methods}

\pn
Most of the simulation-based methods for solving the projected equation use explicitly or implicitly the notion of  temporal difference (TD), which  originated in reinforcement
learning with the works of Samuel [Sam59], [Sam67] on a checkers-playing program. The first TD method is TD($\l$), which can be viewed as an iterative stochastic approximation-type algorithm. The LSTD($\l$) method is based on batch simulation: it first generates a batch of state and cost samples, it approximates the projected equation $\Phi r=\Pi T_\m^{(\l)}(\Phi r)$ using these samples, and then solves the equation directly by matrix inversion. Another TD method is 
LSPE($\l$), which while being more iterative, shares much of the simulation philosophy of LSTD($\l$). 

To describe more specifically the  LSTD($\l$) and 
LSPE($\l$) methods, we first note that the orthogonality condition that characterizes the projection in the projected equation $\Phi r=\Pi T_\m^{(\l)}(\Phi r)$ is
$$\Phi'\Xi\big(\Phi r-T_\m^{(\l)}(\Phi r)\big)=0,\xdef\orthcond{\lab}\eqnum\show{lsp}$$
where $\Xi$ is the diagonal matrix with the vector $\xi$ along the diagonal (see e.g., [Ber07]).
Thus the projected equation \projeqone \ is equivalent to the lower-dimensional equation \orthcond, which can in turn be written in matrix form as\old{\footnote{\dag}
{\ninepoint  For a converse relation note that by premultiplying the equation $C^{(\l)}r=d^{(\l)}$ with $\Phi$, and by using the projection formula $\P J=\Phi(\Phi'\Xi \Phi)^{-1}\Phi'\Xi J$, we obtain the projected equation $\Phi r=\Pi T_\m^{(\l)}(\Phi r)$.}}
$$C^{(\l)}r=d^{(\l)}, \xdef\cdlambdaeq{\lab}\eqnum\show{lsp}$$
with 
$$C^{(\l)}=\Phi'\Xi \big(I-P_\m^{(\l)}\big)\Phi,\qquad d^{(\l)}=\Phi'\Xi g_\m^{(\l)},\xdef\cdlambdaform{\lab}\eqnum\show{lsp}$$
and
$$P_\m^{(\l)}= (1-\l)\sum_{\ell =0}^\infty   \l^\ell  \a^{\ell+1}P_{\m}^{\ell+1}, \qquad  g_\m^{(\l)}=\sum_{\ell=0}^\infty\l^\ell  \a^\ell P_{\m}^\ell g_\m,\eqnum\show{lsp}$$
where $P_\m$ and $g_\m$ are the transition probability matrix and expected single-stage cost vector corresponding to $\m$.
The  LSTD($\l$) and 
LSPE($\l$) methods use simulation-based approximations of $C^{(\l)}$ and $d^{(\l)}$. This is done by simulating a state sequence $(i_0,\ldots,i_t)$ and corresponding transition cost sequence, using the current policy $\m$ (perhaps with exploration enhancement, as discussed earlier). Then after each simulated state $i_\ell$, $\ell=0,\ldots,t$, is generated, estimates $C_{\ell}^{(\l)}$ and $d_{\ell}^{(\l)}$ are obtained using the simulation samples up to time $\ell$, using formulas that we will not give here, as they are not important for our purposes. Such formulas, in various alternative forms,  can be found in several sources, including the textbooks cited earlier. The papers [NeB03], [BeY09], [Yu10a], [Yu10b] discuss the conditions for the convergence $\lim_{\ell\to\infty}C_\ell^{(\l)}= C^{(\l)}$, $\lim_{\ell\to\infty}d_\ell^{(\l)}= d^{(\l)}$ to hold with probability 1. 

The LSTD($\l$) method is based on simple matrix inversion: after the last state $i_t$ of the simulation trajectory is generated, it computes the solution
$$\hat r=\big(C_t^{(\l)}\big)^{-1}d_t^{(\l)}\xdef\iterlamlstd{\lab}\eqnum\show{lsp}$$
 of the corresponding simulation-based approximation to Eq.\ \cdlambdaeq, 
$$C_t^{(\l)}r=d_t^{(\l)},\xdef\approxcdlambdaeq{\lab}\eqnum\show{lsp}$$
and approximates the cost vector $J_\m$ by $\Phi \hat r$. An important point is that $\hat r$ can be obtained regardless of whether $\Pi T^{(\l)}_\m$ is a contraction. It is only required that $C_t^{(\l)}$ is invertible, a much less restrictive condition.

One version of the LSPE($\l$) method
consists of iterative solution of the system \approxcdlambdaeq. It approximates the cost vector $J_\m$ by $\Phi r_{t+1}$,  where $r_{t+1}$ is obtained at the last step of the iteration
$$r_{\ell+1}=r_{\ell}-\g G_{\ell}\bl(C_{\ell}^{(\l)}r_{\ell}-d_{\ell}^{(\l)}\br),\qquad \ell=0,\ldots,t,\xdef\lambdalspe{\lab}\eqnum\show{lsp}$$
where $r_0$ is some initial vector, likely the vector obtained from the preceding policy evaluation, $\g$ is a positive stepsize, $G_{\ell}$ is the matrix
$$G_{\ell}=\lf({1\over \ell}\sum_{m=0}^{\ell-1}\phi(i_m)\phi(i_m)'\ri)^{-1},\qquad \ell=0,\ldots,t,\xdef\gkmatrix{\lab}\eqnum\show{lsp}$$
and as earlier, $\phi(i)'$ denotes the $i$th row of the matrix $\Phi$.
In the original proposal of [BeI96] the stepsize is $\g=1$; convergence of $\Phi r_{t}$ to the fixed point of $\Pi T^{(\l)}_\m$ for this stepsize was shown in [BBN04].
The matrix $G_{\ell}$ is a simulation-based approximation of $(\Phi'\Xi\Phi)^{-1}$ (alternative choices of $G_\ell$ have been discussed recently in  [Ber11b], [Ber11c]). There is also an equivalent implementation of this iteration, which is based on solution of a least squares problem (see Section 4.1).

The choice \gkmatrix\ for $G_{\ell}$ and the use of $\g=1$ are based on a view of the method as an approximation to the projected value iteration method
$$\Phi r_{\ell+1}=\Pi T_\m^{(\l)}(\Phi r_\ell),$$
which after some calculation can be written as
$$\Phi r_{\ell+1}=\Phi \big(r_\ell-(\Phi'\Xi\Phi)^{-1}\big(C^{(\l)}r_\ell-d^{(\l)})\big),$$
or equivalently, since $\Phi$ has full rank, as
$$r_{\ell+1}=r_{\ell}-(\Phi'\Xi\Phi)^{-1}\bl(C^{(\l)}r_{\ell}-d^{(\l)}\br);$$
cf.\ Eq.\ \lambdalspe-\gkmatrix\ with $\g=1$. 

Note that the matrix inversion in Eq.\ \gkmatrix\ is not so onerous, because it can be formed incrementally, with a rank-one correction as each sample becomes available. 
On the other hand, contrary to LSTD($\l$) [and similar to TD($\l$)], the LSPE($\l$) method \lambdalspe-\gkmatrix\ requires that $\Pi T_\m^{(\l)}$ be a contraction for convergence. Indeed if the simulation is performed using the steady-state distribution $\xi_\m$, it can be shown that $\Pi T_\m^{(\l)}$ is a contraction, but if the simulation is performed using a mixture/off-policy distribution \explore\ for the purpose of  exploration-enhancement, the contraction property may be lost and repeated iterations of the form \lambdalspe\ may diverge.

We finally note that in iteration \lambdalspe \ the underlying assumption is 
that we update $r$ as simulation samples are collected and used to form ever improving approximations to $C$ and $d$. An alternative is to use batch simulation, like in LSTD: first simulate to obtain $C_t^{(\l)}$, $d_t^{(\l)}$, and $G_t$, and then solve the system $C_t^{(\l)} r=d_t^{(\l)}$ iteratively rather than through the direct matrix inversion \iterlamlstd, by using any number of iterations of the type \lambdalspe. In fact, we may use {\it only one} iteration, in which case the method takes the form
$$r_{1}=r_{0}-\g G_{t}\bl(C_{t}^{(\l)}r_{0}-d_{t}^{(\l)}\br).\xdef\singlelspe{\lab}\eqnum\show{lsp}$$
A single (or very few) iterations may be sufficient if $\l$ is close to 1, since then the contraction modulus of $\Pi T_{\m}^{(\l)}$ is close to 0 (see e.g., [BeT96], Lemma 6.6, or [Ber07], Prop.\ 6.3.3), so a single iteration with $\Pi T_{\m}^{(\l)}$ is very effective, yielding a vector that is close to its fixed point. We will return to this variant of the method later.

\subsection{Comparison of LSTD($\l$) and LSPE($\l$)}

\pn There has been speculation about the relative merits of LSTD($\l$) and LSPE($\l$). Generally speaking, it is difficult to reach definitive conclusions, as there are several complex factors to consider, such as the length of the simulation sequence $(i_0,\ldots,i_t)$, and the potential near-singularity of $C^{(\l)}$, which affects the  error in the matrix inversion in the LSTD($\l$) formula \iterlamlstd. As an illustration, consider a few different situations:

\nitem{(a)} Assume, as an idealization, that an infinite number of samples is collected. Then both methods yield in the limit the same result, the fixed point of the projected equation 
$J=\Pi T_\m^{(\l)}J.$
 However, in contrast to LSTD($\l$),  in order to guarantee convergence, LSPE($\l$) requires that $\Pi T_\m^{(\l)}$ is a contraction, which interferes with the freedom to do exploration, as discussed earlier.

\nitem{(b)} Assume that $C^{(\l)}$ is invertible, but is nearly singular. Then the matrix inversion in the LSTD($\l$)  formula \iterlamlstd\ may require a very large number of samples to yield a reasonably accurate solution of $C^{(\l)}r=d^{(\l)}$.\footnote{\dag}{\ninepoint It is well-known from fundamental error analyses of linear equation solvers that small errors in a nearly singular matrix $C^{(\l)}$ will cause large errors in the solution of $C^{(\l)}r=d^{(\l)}$. Near-singularity of $C^{(\l)}$ may be due either to the columns of $\Phi$ being nearly linearly dependent or to the matrix $\Xi (I-\a P^{(\l)})$ being nearly singular [cf.\ Eq.\ \cdlambdaform]. Near-linear dependence of the columns of $\Phi$ will not affect the error in the solution of the high-dimensional projected equation, which can be written as $\Phi C^{(\l)}r=\Phi d^{(\l)}$. The reason is that this error depends only on the subspace $S$ and not its representation in terms of the matrix $\Phi$. In particular, if we replace $\Phi$ with a matrix $\Phi B$ where $B$ is an $s\times s$ invertible scaling matrix, the subspace $S$ will be unaffected and the error in the solution of the projected equation will also be unaffected. On the other hand, near singularity of the matrix $I-\a P^{(\l)}$ may affect significantly the error. Note that $I-\a P^{(\l)}$ is nearly singular in the case where $\a$ is very close to 1, or in the corresponding undiscounted case where $\a=1$ and $P$ is substochastic with some eigenvalues very close to 1. Large variations in the size of the diagonal components of $\Xi$ may also affect significantly the error, although this dependence is complicated by the fact that $\Xi$ appears not only in the formula $C^{(\l)}=\Phi'\Xi (I-\a P^{(\l)})\Phi$ but also in the formula $d^{(\l)}=\Phi' \Xi g^{(\l)}$.}
To correct the sensitivity of LSTD($\l$) to simulation noise, it may be necessary to turn it into an iterative method through some form of regularization, which then brings it close to a form of LSPE($\l$) (see [Ber09], [WPB09], [Ber11a], [Ber11b], [Ber11c] for such regularization methods and their connection to LSPE). Of course, the situation becomes even more complex if $C^{(\l)}$ is singular, perhaps due to inadvertent rank deficiency of $\Phi$ (see [WaB11a], [WaB11b] for a discussion of this possibility).

\nitem{(c)} When LSTD($\l$) and LSPE($\l$) are embedded within a PI framework, the number of samples collected using any one policy is often relatively small. Then the behavior of the two methods becomes very complicated, and it is hard to reach any kind of reliable conclusion [Ber10]. Computational studies indicate that  LSPE($\l$) being an iterative method, is less sensitive to the matrix inversion errors that afflict  LSTD($\l$) in the presence of high simulation noise.

\smskip

\pn The preceding discussion is also relevant to the implementations of $\l$-PI to be discussed in the next section, since these implementations bear strong relations to both LSTD($\l$) and LSPE($\l$).

\vskip-1.5pc

\section{Lambda-Policy Iteration with Cost Function Approximation}
\vskip-0.1pc

\pn 
We saw in Section 2 that the policy evaluation portion of $\l$-PI,
$$J_{k+1}=T_{\m_{k+1}}^{(\l)}J_{k},\xdef\lpiiterate{\lab}\eqnum\show{lsp}$$
[cf.\ Eq.\ \lamdaeval] can be implemented in two ways:

\nitem{(1)} By computing $T_{\m_{k+1}}^{(\l)}J_{k}$.

\nitem{(2)} By finding the fixed point of the mapping 
$W_k$
[cf.\ Eq.\ \mapm] through solution of the equation
$$J=W_kJ,\xdef\beleqstopa{\lab}\eqnum\show{lsp}$$
which can be viewed as Bellman's equation associated with the current policy for the two equivalent DP problems discussed in Section 2 [cf.\ Eq.\ \beleqstop]: a $\l\a$-discounted problem and a stopping problem.
\smskip

Let us now consider approximation of $\l$-PI on the subspace $S=\{\Phi r\mid r\in\re^s\}$. A natural possibility is to introduce projection in the preceding approaches. In particular, we may approximate the $\l$-PI iterate $J_{k+1}$ of Eq.\ \lpiiterate\ by $\Phi r_{k+1}$ in three ways:

\nitem{(a)} By using a single projected value iteration for the original $\a$-discounted problem,
$$\Phi r_{k+1}=\Pi T_{\m_{k+1}}^{(\l)}(\Phi r_k).\xdef\lambdaprojpi{\lab}\eqnum\show{lsp}$$
This is the original proposal of [BeI96].  It is the variant of the LSPE($\l$) method \lambdalspe-\gkmatrix, which involves just the last iteration. 

\nitem{(b)} By solving a projected version of Eq.\ \beleqstopa, viewing it as Bellman's equation for  the $\l\a$-discounted problem of Section 2,
and setting $r_{k+1}$ equal to its solution. This is the proposal of [ThS10a], and implements by simulation the solution of this projected equation, essentially by applying LSTD(0) to Bellman's equation for the $\l\a$-discounted problem formulated in Section 2. 

\nitem{(c)} By solving a projected version of Eq.\ \beleqstopa, viewing it as Bellman's equation for  the  stopping problem formulated in Section 2, and setting $r_{k+1}$ equal to its solution. 

In the following three subsections, we will describe three alternative implementations of $\l$-PI corresponding to the possibilities (a)-(c) above. 
Of course when linear cost function approximation of the form $\Phi r_k$ is used to represent $J_k$, the $\l$-PI method need not converge, and the cost vectors $J_{\m_k}$ of the generated policies typically oscillate within some suboptimality threshold from $J^*$. We do not address this issue, but we note that related error bounds, which also apply to other forms of optimistic PI are given by  Bertsekas and Yu [BeY10a], Thiery and Scherrer [ThS10b], and Scherrer [Sch11].

\subsection{The LSPE($\l$) Implementation}

\pn  A variant of the LSPE($\l$) method \lambdalspe-\gkmatrix\ is to form batches of simulation samples and perform iteration \lambdalspe\ at the end of each batch. In an extreme case, we treat the entire simulation trajectory $(i_0,\ldots,i_t)$ as a single simulation batch, and we perform a {\it single} iteration \lambdalspe, for $\ell=t$, yielding the method 
$$r_{k+1}=r_k-G_t\bl(C_{t}^{(\l)}r_k-d_{t}^{(\l)}\br),\xdef\lspebatch{\lab}\eqnum\show{lsp}$$
where $\Phi r_k$ is the approximate evaluation of the cost vector of the preceding policy $\m_k$ [cf.\ Eq.\ \singlelspe].
As  $t\to\infty$ and the simulation becomes exact in the limit, i.e.,  
$$\lim_{t\to\infty}C_t^{(\l)}= C^{(\l)},\qquad \lim_{t\to\infty}d_t^{(\l)}= d^{(\l)},$$
and if $G_t$ is given by the formula \gkmatrix, it can be verified that
$$\Phi r_{k+1}\to \Pi T_{\m_{k+1}}^{(\l)}(\Phi r_k).\xdef\lspeconv{\lab}\eqnum\show{lsp}$$
Thus the method \lspebatch\ with $G_t$ given by Eq.\ \gkmatrix\ can be viewed as a simulation-based implementation of Eq.\ \lambdaprojpi, the projected version of $\l$-PI, which becomes exact in the limit as $t\to\infty$. In practice of course $t$ is finite, and one may consider variants of the method, whereby multiple iterations of the form \lspebatch\ are performed, with each iteration using additional simulation samples.

We note a mathematically equivalent description of this method, which is given in terms of a least-squares optimization (see [Ber07], Section 6.3.3 for a more detailed textbook account): we set
$$r_{k+1}=\arg\min_{r\in\re^s}\sum_{\ell=0}^t\lf(\phi(i_\ell)'r-\phi(i_\ell)'r_k-\sum_{m=\ell}^t(\l\a)^{m-\ell}q(i_m,i_{m+1})\ri)^2,\xdef\lpspelso{\lab}\eqnum\show{lsp}$$
where $q(i_m,i_{m+1})$ is the temporal difference
$$q(i_m,i_{m+1})=g\big(i_m,\m_{k+1}(i_m),i_{m+1}\big)+\a\phi(i_{m+1})'r_k-\phi(i_{m})'r_k,\qquad m=0,\ldots,t.\xdef\lpspelst{\lab}\eqnum\show{lsp}$$
In fact this is how the method was originally described in [BeI96] and [BeT96].

A positive aspect of this method is that it approximates directly $\Pi T_{\m_{k+1}}^{(\l)}(\Phi r_k)$, so it is not subject to bias in the evaluation of the fixed point of $W_k$; cf.\ Eq.\ \lspeconv. However, in the form given here, the method does not address the issue of exploration. Despite this fact, this implementation [in the form \lpspelso] has been successful in several challenging computational studies, including the one involving the game of tetris in the original paper [BeI96] and some followup works, and a recent one by Foderaro et.\ al.\ [FRF11] involving the game of pac-man, a benchmark problem of pursuit-evasion. 

\subsection{$\l$-PI(0) - An Implementation Based on a Discounted MDP}

\pn This implementation, suggested and tested by Thiery and Scherrer [ThS10a], [ThS10b], is based on the fixed point property of $J_{k+1}$ [cf.\ Prop.\ \proplpolt (b)]. It produces an approximation $\Phi r_{k+1}$ to $J_{k+1}$  within the subspace $S$, by solving the projected equation
$$\Phi r=\Pi W_k(\Phi r),$$
with $W_k$ given by 
$$W_kJ=(1-\l)T_{\m_{k+1}}(\Phi r_k)+\l
T_{\m_{k+1}}J,\qquad J\in \rn.\xdef\mapmapprox{\lab}\eqnum\show{lsp}$$
We may find the solution $r_{k+1}$ of this equation by using an LSTD(0)-like simulation approach. In particular, $r_{k+1}$ satisfies the orthogonality condition
$$Cr=d(k),$$
where
$$C=\Phi'\Xi(I-\l\a P_{\m_{k+1}})\Phi,\qquad 
d(k)=\Phi'\Xi \big(g_{\m_{k+1}}+(1-\l)\a P_{\m_{k+1}}\Phi r_k\big),$$
so that
$$r_{k+1}=C^{-1}d(k).\xdef \scherrerit{\lab}\eqnum\show{lsp}$$
We refer to this method as $\l$-PI(0) to distinguish it notationally from the method of the next subsection (the name LS$\l$PI was introduced for this method in [ThS10a]). 

In a simulation-based implementation, the matrix $C$ and the vector $d(k)$ are approximated by estimates $C_t$ and $d_t(k)$. Thus this method does not require that $\Pi T_{\m_{k+1}}^{(\l)}$ is a contraction, and like LSTD, it can deal well with the issue of exploration.  The simulation samples need not depend on the policy $\m_{k+1}$ being evaluated, so they can be generated only once within a PI process. On the other hand the objective of the implementation is to approximate the next iterate of $\l$-PI, i.e., $T^{(\l)}_{\m_{k+1}}(\Phi r_k)$, and it is not clear that it is doing this well. To see this, suppose that the iteration \scherrerit, or equivalently $\Phi r_{k+1}=\Pi W_k(\Phi r_k)$, is repeated an infinite number of times so it converges to a limit $\ol r$, which must satisfy $\Phi \ol r=\Pi W_k(\Phi \ol r)$. Then using Eq.\ \mapmapprox, we have
$$\Phi \ol r=(1-\l)\Pi T_{\m_{k+1}}(\Phi \ol r)+\l
\Pi T_{\m_{k+1}}(\Phi \ol r),$$
 which shows that $\Phi \ol r=\Pi T_{\m_{k+1}}(\Phi \ol r)$. Thus $\l$-PI(0) aims at $\ol r$, which is the limit of TD(0)  independent of the value of $\l$. Indeed as $\l\to1$, $\Pi W_k$ tends to $\Pi T_{\m_{k+1}}$ [cf.\ Eq.\ \mapmapprox], so its fixed point $\Phi r_{k+1}$ tends to the fixed point of  $\Pi T_{\m_{k+1}}$, i.e., the limit of TD(0). It follows that while this implementation deals well with the issue of exploration, it may be subject to significant bias-related error. 
 
 \subsection{$\l$-PI(1) - An Implementation Based on a Stopping Problem}

\pn The third implementation is based on the property mentioned in Section 2: the fixed point equation $J=W_kJ$ [or equivalently, Eq.\ \beleqstop] is Bellman's equation for the policy $\m_{k+1}$ in the context of a stopping problem. Here there is  an artificial termination state 0, and for all states $j$, there is probability $1-\l$ that a transition to $j$ will be followed by an immediate transition to state 0, with cost $\a J_k(j)$, cf.\ Eq.\ \beleqstop. Note that if $\l$ is not too close to 1, the trajectories of this problem tend to be short, and in fact if $\l=0$ all trajectories consist of a single transition.

To compute an approximation $\Phi r_{k+1}$ to the fixed point of $W_k$ by using the stopping problem, we may use any policy evaluation algorithm with cost function approximation over the subspace $S=\{\Phi r\mid r\in\re^s\}$.  An interesting choice is to use the LSPE(1) method, which consists of a least squares fit of $\Phi r$ to the simulated costs of the trajectories of the stopping problem whose Bellman equation mapping is $W_k$.  The use of LSPE(1) not only involves minimum bias relative to all  LSPE($\nu$) methods with $\nu\in[0,1]$, but also leads to a simple least squares implementation.

To this end, we introduce a simulation procedure, called {\it geometric sampling\/}, which departs from the single infinitely long simulation trajectory format of the implementation of Section 4.1, and has the following characteristics:

\nitem{(a)} It uses multiple relatively short simulation trajectories. 
\nitem{(b)} The initial state of each trajectory is chosen essentially as desired, thereby allowing flexibility to generate a richer mixture of state visits. 
\nitem{(c)} The length of each trajectory is random and is determined by a $\l$-dependent geometric distribution [a probability $(1-\l)\l^\ell$ that the number of transitions is $\ell+1$]. \old{This allows the simultaneous simulation-based approximation of all the terms $(1-\l)\l^\ell T^{\ell+1}J$ in the series expansion of $T^{(\l)}J$.}
\smskip

In particular, given the current representation $\Phi r_k$ of $J_k$ and the current policy $\m_{k+1}$, we update the parameter vector from $r_k$ to $r_{k+1}$ after generating $t$ simulated trajectories. The states of a trajectory are generated according to the transition probabilities $p_{ij}\big(\m_{k+1}(i)\big)$, the transition cost is discounted by an additional factor $\a$ with each transition, and following each transition to a state $j$, the trajectory is terminated with probability $1-\l$ and with an extra cost $\a \phi(i)'r_k$.
Once a trajectory is terminated, an initial state for the next trajectory is
chosen according to a fixed probability distribution
$\zeta_0=\bl(\zeta_0(1),\ldots,\zeta_0(n)\br)$,
and the process is repeated. Note that the sequence of restart states need not depend on the policy being evaluated, so that it can be simulated only once within a PI process. Of course, the simulated trajectories have to be recalculated for each new policy. The details are as follows.

Let the $m$th trajectory, $m=1,\ldots,t$,  have the form
$(i_{0,m},i_{1,m},\ldots,i_{N_m,m})$, where $i_{0,m}$ is the initial state, and $i_{N_m,m}$ is the state at which the trajectory is completed (the last state prior to termination). For each state $i_{\ell,m}$, $\ell=0,\ldots,N_m-1$, of the $m$th trajectory, the simulated cost is
$$c_{\ell, m}(r_k)=\a^{N_m-\ell}\phi(i_{N_m,m})'r_k+\sum_{q=\ell}^{N_m-1} \a^{q-\ell} g(i_{q,m},u_{q,m},i_{q+1,m}),\xdef \costsamples{\lab}\eqnum\show{lsp}$$
where
$$u_{q,m}=\m_{k+1}(i_{q,m}),\qquad m=1,\ldots,t,\ q=0,\ldots,N_m-1.$$
Once the costs $c_{\ell, m}(r_k)$ are computed for all states $i_{\ell, m}$ of the $m$th trajectory and all trajectories $m=1,\ldots,t$, the vector $r_{k+1}$ is obtained by a least squares fit of these costs:
$$r_{k+1}=\arg\min_{r\in\re^s}\sum_{m=1}^t\sum_{\ell=0}^{N_m-1}\big(\phi(i_{\ell,m})'r-c_{\ell,m}(r_k)\big)^2,\eqnum\show{lsp}$$
cf.\ Eqs.\ \lpspelso-\lpspelst. Equivalently, we can write the  solution of the least squares problem explicitly as
$$r_{k+1}=\lf(\sum_{m=1}^t\sum_{\ell=0}^{N_m-1}\phi(i_{\ell, m})\phi(i_{\ell, m})'\ri)^{-1} \sum_{m=1}^t\sum_{\ell=0}^{N_m-1}\phi(i_{\ell, m})c_{\ell, m}(r_k).\xdef \lscostsamples{\lab}\eqnum\show{lsp}$$
We refer to the resulting implementation as $\l$-PI(1).

Note the extreme special case when $\l=0$. Then all the simulated trajectories consist of a single transition, and there is a restart at every transition. This means that the simulation samples are from states that are  generated independently according to the restart distribution $\zeta_0$.

\subsubsection{Convergence of the Simulation Process}

\pn We will now show that in the limit, as $t\to\infty$, the vector $r_{k+1}$ of Eq.\ \lscostsamples\ satisfies
$$\Phi r_{k+1}=\hat \Pi T_{\m_{k+1}}^{(\l)}(\Phi r_k),\xdef \lscostlim{\lab}\eqnum\show{lsp}$$
where $\hat \Pi$ denotes projection with respect to the weighted sup-norm $\|\cdot\|_\zeta$ with weight vector $\zeta=\big(\zeta(1),\ldots,\zeta(n)\big)$, where 
$$\zeta(i)={\hat\zeta(i)\over \sum_{j=1}^n\hat\zeta(j)},\qquad i=1,\ldots,n,\eqnum\show{lsp}$$
and 
$$\hat \zeta(i)=\sum_{\ell=0}^\infty \zeta_\ell(i),$$ 
with $\zeta_\ell(i)$ being the probability of the state being $i$ after $\ell$ transitions of a randomly chosen simulation trajectory. This is the underlying norm in TD methods such as LSTD, LSPE, and TD, as applied to SSP problems (see [BeT96], Section 6.3.4). Note that $\zeta (i)$ is the long-term occupancy probability of state $i$ during the simulation process. We assume that the restart distribution $\zeta_0$ is chosen so that $\zeta(i)>0$ for all $i=1,\ldots,n$, implying that $\|\cdot\|_\zeta$ is a legitimate norm [this is guaranteed if we require that $\zeta_0(i)>0$ for all $i$].

Indeed, let us 
view $T^{\ell +1}_{\m_{k+1}}J$ as the vector of total discounted costs over a
horizon of 
$(\ell +1)$ stages with the terminal cost function being
$J$, and write
$$T^{\ell +1}_{\m_{k+1}}J=\a^{\ell +1}P^{\ell +1}_{\m_{k+1}}J+\sum_{q=0}^\ell \a^q P_{\m_{k+1}}^q g_{\m_{k+1}},\old{\eqnum\show{lsp}}$$ 
where $P_{\m_{k+1}}$ and $g_{\m_{k+1}}$ are the transition probability matrix and cost vector, respectively, under $\m_{k+1}$. As a result the vector $T_{\m_{k+1}}^{(\l)}J=(1-\l)\sum_{\ell=0}^\infty\l^ \ell T_{\m_{k+1}}^{\ell  +1}J$ can be expressed as
$$\big(T_{\m_{k+1}}^{(\l)}J \big)(i)=\sum_{\ell=0}^\infty(1-\l)\l^\ell E\lf\{\a^{\ell +1} J(i_{\ell +1})+\sum_{q=0}^{\ell} \a^q
g\big(i_{q},\m_{k+1}(i_q), i_{q+1}\big)\ \Big|\ i_0=i\ri\},\qquad i=1,\ldots,n.$$
Thus $\big(T_{\m_{k+1}}^{(\l)}J \big)(i)$ may be viewed as the expected value of the $(\ell+1)$-stages cost of policy  $\m_{k+1}$ starting at state $i$, with {\it the number of stages being random and geometrically distributed with parameter $\lambda$}  [probability of $\kappa+1$ transitions is $(1-\l)\l^ \kappa $, $\kappa =0,1,\ldots$]. It follows that the cost samples $c_{\ell, m}(r_k)$  of Eq.\ \costsamples, produced by the simulation process described earlier, can be used to estimate $\big(T_{\m_{k+1}}^{(\l)}(\Phi r_k) \big)(i)$ for all $i$ by Monte Carlo averaging. The estimation formula is
$$D_t(i)={1\over \sum_{m=1}^t\sum_{\ell=0}^{N_m-1}
\d(i_{\ell,m}=i)}\cdot \sum_{m=1}^t\sum_{\ell=0}^{N_m-1}
\d(i_{\ell,m}=i)c_{\ell,m}(r_k),\xdef\costestti{\lab}\eqnum\show{lsp}$$
where $\d(i_{\ell,m}=i)=1$ if $i_{\ell,m}=i$ and $\d(i_{\ell,m}=i)=0$ otherwise, and we have
$$\big(T_{\m_{k+1}}^{(\l)}(\Phi r_k) \big)(i)=\lim_{t\to\infty}D_t(i),\qquad i=1,\ldots,n,$$
(see also the discussion on the consistency of Monte Carlo simulation for policy evaluation in [BeT96], Section 5.2). 

Let us now compare the $\l$-PI iteration  \lscostlim\ with the simulation-based implementation \lscostsamples. Using the definition of projection, Eq.\ \lscostlim\ can be written as
$$r_{k+1}=\arg\min_{r\in\re^s}\sum_{i=1}^n\zeta(i)\Big(\phi(i)'r-\big(T_{\m_{k+1}}^{(\l)}(\Phi r_k) \big)(i)\Big)^2,$$
or equivalently
$$r_{k+1}=\lf(\sum_{i=1}^n\zeta(i)\phi(i)\phi(i)'\ri)^{-1}\sum_{i=1}^n\zeta(i)\phi(i)\big(T_{\m_{k+1}}^{(\l)}(\Phi r_k) \big)(i).\xdef\exactproj{\lab}\eqnum\show{lsp}$$
Let $\tl \zeta(i)$ be the empirical relative frequency of state $i$ during the simulation, given by
$$\tl \zeta(i)={1\over N_1+\cdots+N_t}\sum_{m=1}^t\sum_{\ell=0}^{N_m-1}
\d(i_{\ell,m}=i).\xdef\freqestti{\lab}\eqnum\show{lsp}$$
Then the simulation-based estimate \lscostsamples\ can be written as
$$\eqalign{r_{k+1}&=\lf(\sum_{m=1}^t\sum_{\ell=0}^{N_m-1}\phi(i_{\ell, m})\phi(i_{\ell, m})'\ri)^{-1} \sum_{m=1}^t\sum_{\ell=0}^{N_m-1}\phi(i_{\ell, m})c_{\ell, m}(r_k)\cr
&=\lf(\sum_{i=1}^n\sum_{m=1}^t\sum_{\ell=0}^{N_m-1}\d(i_{\ell,m}=i)\phi(i)\phi(i)'\ri)^{-1}\sum_{i=1}^n \sum_{m=1}^t\sum_{\ell=0}^{N_m-1}\d(i_{\ell,m}=i)\phi(i)c_{\ell, m}(r_k)\cr
&=\lf(\sum_{i=1}^n\tl \zeta(i)\phi(i)\phi(i)'\ri)^{-1}\sum_{i=1}^n{1\over N_1+\cdots+N_t}\cdot \phi(i) \cdot \sum_{m=1}^t\sum_{\ell=0}^{N_m-1}\d(i_{\ell,m}=i)c_{\ell, m}(r_k)\cr
&=\lf(\sum_{i=1}^n\tl \zeta(i)\phi(i)\phi(i)'\ri)^{-1}\sum_{i=1}^n {\sum_{m=1}^t\sum_{\ell=0}^{N_m-1}
\d(i_{\ell,m}=i)\over N_1+\cdots+N_t}\cdot \phi(i)  \cdot \cr
&\ \ \  \ \ \ \ \ \ \ \ \  \ \ \ \ \ \ \ \ \  \ \ \ \ \ \ \ \ \  \ \ \ \ \ \ \ \cdot {1\over \sum_{m=1}^t\sum_{\ell=0}^{N_m-1}
\d(i_{\ell,m}=i)} \cdot \sum_{m=1}^t\sum_{\ell=0}^{N_m-1}\d(i_{\ell,m}=i)c_{\ell, m}(r_k)\cr}$$
and finally, using Eqs.\ \costestti\ and \freqestti,
$$r_{k+1}=\lf(\sum_{i=1}^n\tl \zeta(i)\phi(i)\phi(i)'\ri)^{-1}\sum_{i=1}^n\tl \zeta(i)\phi(i)D_t(i).\xdef\approxproj{\lab}\eqnum\show{lsp}$$
We can now compare the $\l$-PI iteration  \exactproj\ and the simulation-based implementation \approxproj. Since $\big(T_{\m_{k+1}}^{(\l)}(\Phi r_k) \big)(i)=\lim_{t\to\infty}D_t(i)$ and $\zeta(i)=\lim_{t\to\infty}\tl \zeta(i)$, we see that these two iterations asymptotically coincide.

The expression \approxproj\ provides some insight on how $\l$-PI(1) approximates the $\l$-PI iteration \exactproj\ [or equivalently $\Phi r_{k+1}= \hat\Pi T^{(\l)}_{\m_{k+1}}(\Phi r_k)$; cf.\ Eq.\ \lscostlim]. Generally the simulation process of $\l$-PI(1) (many short trajectories) involves more noise than the simulation process of the other implementations (a single long trajectory), because the length of each simulation trajectory is random (exponentially distributed). This can be seen from iteration \approxproj, which involves considerable simulation noise due to the presence of $\tl \zeta(i)$ and $D_t(i)$. However, we will argue that from a practical point of view much of this noise does not play a significant role. To see this, first note that the deviation of $\tl \zeta(i)$ from $\zeta(i)$, is not important since $\tl \zeta(i)$ simply redefines the projection norm. Next note that $D_t(i)$ can be written as
$$D_t(i)=\sum_{\ell=0}^\infty \tl f_\ell(i)\tl E_\ell(i),\xdef\empirical{\lab}\eqnum\show{lsp}$$ 
where $\tl f_\ell(i)$ and $\tl E_\ell(i)$ are the following empirical averages over the entire simulation process:
\nitem{(a)} $\tl f_\ell(i)$ is the empirical relative frequency of cost samples that start at state $i$, and correspond to trajectories consisting of $\ell+1$ transitions. As $t\to\infty$ it converges to $(1-\l)\l^{\ell}$ based on the way the simulation is structured.
\nitem{(b)} $\tl E_\ell(i)$ is the Monte Carlo estimate of the cost of trajectories that start at state $i$, consist of $\ell+1$ transitions, and have terminal cost vector $\Phi r_k$. As $t\to\infty$ it converges to $T^{\ell+1}_{\m_{k+1}}(\Phi r_k)(i)$.
\smskip
\pn While both $\tl f_\ell(i)$ and $\tl E_\ell(i)$ contribute to the variance of $D_t(i)$, only $\tl E_\ell(i)$ has practical significance. To see this note that based on Eq.\ \empirical, $D_t(i)$ can also be viewed as an estimate of 
$$\tl T_{\m_{k+1}}(\Phi r_k)(i)=\sum_{\ell=0}^\infty \tl f_\ell(i)T^{\ell+1}_{\m_{k+1}}(\Phi r_k)(i).\xdef\empiricalo{\lab}\eqnum\show{lsp}$$
Thus iteration \approxproj\ may also be viewed as a simulation-based implementation of the optimistic PI method
$$\Phi r_{k+1}=\tl \Pi \tl T_{\m_{k+1}}(\Phi r_k),$$
where $\tl \Pi$ is projection with respect to the weighted sup-norm defined by $\tl \zeta$. From a practical point of view, this iteration and the $\l$-PI iteration $\Phi r_{k+1}= \hat \Pi T^{(\l)}_{\m_{k+1}}(\Phi r_k)$ perform similarly: there is only a difference in the projection norm ($\tl \Pi$ rather than $\hat \Pi$), and a difference in the weights of the terms $T^{\ell+1}_{\m_{k+1}}$ [$\tl f_\ell(i)$ rather than $(1-\l)\l^\ell$]; compare $\tl T_{\m_{k+1}}(\Phi r_k)(i)$ as given by Eq.\ \empiricalo\ with
$$T^{(\l)}_{\m_{k+1}}(\Phi r_k)(i)=\sum_{\ell=0}^\infty (1-\l)\l^\ell T^{\ell+1}_{\m_{k+1}}(\Phi r_k)(i),$$
the definition of $T^{(\l)}_{\m_{k+1}}$. Neither difference should affect significantly the quality of the obtained approximation $\Phi r_{k+1}$.

In conclusion, with the $\l$-PI(1) implementation \costsamples-\lscostsamples, as $t\to\infty$, we obtain in the limit the $\l$-PI iteration Eq.\ \lscostlim, with comparable performance degradation due to simulation noise as for the LSPE($\l$) implementation of Section 4.1.
A key characteristic of the implementation is that it deals with the issue of exploration flexibly and effectively. Since a trajectory of the stopping problem is completed at each transition with the potentially large probability $1-\l$, a restart with a new initial state $i_0$ is frequent and the length of each of the simulated trajectories is relatively small. The restart mechanism can be used as a ``natural" form of exploration, by choosing appropriately the restart distribution $\zeta_0$ so that $\zeta(i)$ reflects a ``substantial" weight for all states $i$. Thus $\l$-PI(1) is like LSPE($\l$) (Section 4.1), but with built-in exploration enhancement. Compared to  $\l$-PI(0) (Section 4.2) it involves reduced bias since it aims to find the limit point of TD($\l$), not TD(0). In particular, as $\l\to1$, it produces an evaluation $\Phi r_{k+1}$ that tends to the fixed point of TD(1), i.e., the projection $\hat \Pi J_{\m_{k+1}}$.

\subsection{Comparison with Alternative Approximate PI Methods}

\pn The preceding $\l$-PI implementations are in direct competition with approximate PI methods that use LSTD($\l$) for policy evaluation. A popular method, often referred to as LSPI (Lagoudakis and Parr [LaP03]), can be simply described as approximate PI combined with LSTD(0) for policy evaluation. The LSPI and $\l$-PI(0) methods have been compared in [ThS10a] in terms of four characteristics.

\nitem{(a)} {\it Bias\/}: Both methods are subject to qualitatively similar bias [they aim to find the limit point of TD(0)].

\nitem{(b)} {\it Sample efficiency\/}: Both methods can reuse the same set of sample state trajectories over all policies. (In the model-free case where Q-factors are approximated, again the set of sample state-control trajectories is reusable.)

\nitem{(c)} {\it Exploration\/}: Both methods provide the same options for exploration, since the validity of these methods does not depend on whether the simulation trajectories are obtained by using the current policy [in fact these trajectories are reusable as per (b) above].

\nitem{(d)} {\it Optimistic operation\/}: Since $\l$-PI(0) has an iterative character ($r_{k+1}$ depends on $r_k$), it is less susceptible to simulation noise and has an advantage over LSPI in the case where the number of samples per policy is low. Indeed this assertion is made by Thiery and Scherrer [ThS10a] based on experimentation, who also found that the effect of the choice of $\l$ is more pronounced in this case.

\smskip

\pn Note that (b) and (c) above are the advantages of LSPI and $\l$-PI(0) over the LSPE($\l$) implementation of Section 4.1 (which in turn  involves less bias because of the use of $\l>0$, and also has an optimistic character). 

Let us now compare 
$\l$-PI(1) with LSPI and $\l$-PI(0) in terms of the characteristics (a)-(d) above. It has better bias characteristics as noted earlier. It has worse sample efficiency as it cannot reuse simulation trajectories (it can only reuse the restart state sequence). It deals with exploration about as well, thanks to the restart mechanism of the SSP formulation. Finally, like $\l$-PI(0), $\l$-PI(1) has an optimistic character, and has a similar advantage over LSPI in this regard, cf.\ (d) above.

\subsection{Exploration-Enhanced LSTD($\l$) with Geometric Sampling}

\pn The  geometric sampling idea underlying the $\l$-PI(1) implementation  of Eqs.\ \costsamples-\lscostsamples\ may also be modified to obtain an exploration-enhanced version of LSTD($\l$).  In particular, we use the same simulation procedure, and in analogy to Eq.\ \costsamples\ we define
$$c_{\ell, m}(r)=\a^{N_m-\ell}\phi(i_{N_m,m})'r+\sum_{q=\ell}^{N_m-1} \a^{q-\ell} g(i_{q,m},u_{q,m},i_{q+1,m}).$$
We then obtain an approximation $\Phi \hat r$ to the solution of the projected equation 
$$\Phi r=\hat \Pi T_{\m_{k+1}}^{(\l)}(\Phi r),$$
[cf.\ Eq.\ \lscostlim] by finding $\hat r$ such that
$$\hat r=\arg\min_{r\in\re^s}\sum_{m=1}^t\sum_{\ell=0}^{N_m-1}\big(\phi(i_{\ell,m})'r-c_{\ell,m}(\hat r)\big)^2.\xdef \costsamplestz{\lab}\eqnum\show{lsp}$$
By writing the optimality condition 
$$\sum_{m=1}^t\sum_{\ell=0}^{N_m-1}\phi(i_{\ell,m})\big(\phi(i_{\ell,m})'\hat r-c_{\ell,m}(\hat r)\big)=0$$
for the least squares minimization in Eq.\ \costsamplestz\ and solving for $\hat r$, we obtain the following implementation of LSTD($\l$):
$$\hat r=\hat C^{-1}\hat d,\xdef \costsamplesto{\lab}\eqnum\show{lsp}$$
where
$$\hat C=\sum_{m=1}^t\sum_{\ell=0}^{N_m-1}\phi(i_{\ell,m})\big(\phi(i_{\ell,m})-\a^{N_m-\ell}\phi(i_{N_m,m})\big)',\xdef \costsamplestt{\lab}\eqnum\show{lsp}$$
and
$$\hat d=\sum_{m=1}^t\sum_{\ell=0}^{N_m-1}\phi(i_{\ell,m})\sum_{q=\ell}^{N_m-1} \a^{q-\ell} g(i_{q,m},u_{q,m},i_{q+1,m}).\xdef \costsamplestth{\lab}\eqnum\show{lsp}$$
For a large number of trajectories $t$, the exploration-enhanced LSTD($\l$) method \costsamplestz\ [or equivalently \costsamplesto-\costsamplestth] and $\l$-PI(1) [cf.\ Eq.\  \lscostsamples]  yield similar results, particularly when $\l\approx 1$. However, $\l$-PI(1) has an iterative character ($r_{k+1}$ depends on $r_k$), so it is reasonable to expect that it is less susceptible to simulation noise in an optimistic PI setting where the number of samples per policy is low.

As an example, when $\l=0$, all the simulation trajectories consist of a single transition, so $N_m=1$ for all $m=1,\ldots,t$. Then, using Eqs.\ \costsamplestt\ and \costsamplestth, the equation $\hat Cr=\hat d$ becomes
$$\sum_{m=1}^t\phi(i_{0,m})\big(\phi(i_{0,m})-\a\phi(i_{1,m})\big)'r=\sum_{m=1}^t\phi(i_{0,m})g(i_{0,m},u_{0,m},i_{1,m}).$$
It yields the same vector $\hat r=\hat C^{-1}\hat d$ as the LSTD(0) method that simulates $t$ independent transitions according to the restart distribution $\zeta_0$, rather than simulating a single long trajectory. In fact this is the policy evaluation process in the LSPI method mentioned in Section 4.4.  The geometric sampling procedure described here allows exploration-enhancement for any $\l$.

\vskip  -1.5pc
\section{Conclusions}
\mark{Conclusions}
\vskip  -1.5pc

\pn We discussed a few implementations of $\l$-PI with linear cost function approximation, which have different strengths and weaknesses with respect to dealing with the critical issues of bias and exploration. Out of the three implementations, the one of Section 4.3, $\l$-PI(1), is new and seems capable of dealing well with both issues, although it has worse sample complexity than the $\l$-PI(0) implementation of Section 4.2. 

On the other hand, our discussion has been somewhat speculative, and our assessments, while relying on past computational experience, still require supportive experimentation. Moreover, the $\l$-PI implementations should be compared to other approximate PI methods based on projected equations, such as the exploration-enhanced LSTD($\l$) method for policy evaluation, discussed in Section 3, and the LSPI method discussed in Section 4.4. A computational comparison of $\l$-PI(0) with this latter method is given in [ThS10a], and a similar comparison with $\l$-PI(1) would be desirable.

Fundamentally, $\l$-PI(1) is based on geometric sampling, a new simulation idea for $\l$-methods that uses multiple short trajectories with exploration-enhanced restart, rather than a single infinitely long trajectory. This idea can also be applied to LSTD($\l$), thereby obtaining a new exploration-enhanced version of this method, which has been described in Section 4.5.

\section{References}

\mark{References}
\vskip  -0.5pc

\def\refer{\vskip0pt\par\noindent}
\def\ref{\vskip0pt\par\noindent}

\ninepoint

\refer[BBD10] Busoniu, L., Babuska, R., De Schutter, B., and Ernst, D.,  2010.\ Reinforcement Learning and Dynamic Programming Using Function Approximators, CRC Press, N.\ Y.

\refer[BBN04] Bertsekas, D.\ P., Borkar, V.\ S., and Nedi\'c, A., 2004.\ ``Improved Temporal Difference Methods
with Linear Function Approximation," in Learning and Approximate Dynamic Programming, by J.\ Si, A.\ Barto,
W.\ Powell, and D.\ Wunsch (Eds.), IEEE Press, N.\ Y.

\old{\refer[BBS95] Barto, A.\  G., Bradtke, S.\ J., and Singh, S.\ P., 1995.
``Real-Time Learning and Control Using Asynchronous Dynamic Programming," Artificial Intelligence, 
Vol.\ 72, pp.\ 81-138.}

\refer[BSA83] Barto, A.\ G., Sutton, R.\ S., and Anderson, C.\ W., 1983.\ 
``Neuronlike Elements that Can Solve Difficult Learning Control
Problems,'' IEEE Trans.\ on Systems, Man, and Cybernetics,
Vol.\ 13, pp.\ 835-846.

\refer[BeI96]
Bertsekas, D.\ P., and Ioffe, S., 1996.\
``Temporal Differences-Based Policy Iteration and
Applications in Neuro-Dynamic Programming,"
Lab.\ for Info.\ and Decision Systems Report
LIDS-P-2349, MIT.

\refer[BeT96]
Bertsekas, D.\ P., and Tsitsiklis, J.\ N.,
1996.\
Neuro-Dynamic Programming, Athena Scientific, Belmont, MA.

\refer[BeY09] Bertsekas, D.\ P., and Yu, H., 2009.\ ``Projected Equation Methods for Approximate Solution of Large Linear Systems,"  Journal of Computational and Applied Mathematics, Vol.\ 227, pp.\ 27-50.

\refer[BeY10a] Bertsekas, D.\ P., and Yu, H., 2010.\ ``Q-Learning and Enhanced Policy Iteration in Discounted 
Dynamic Programming,"  Lab.\ for Information and Decision Systems Report LIDS-P-2831, MIT.

\refer[BeY10b] Bertsekas, D.\ P., and Yu, H., 2010.\ ``Asynchronous Distributed Policy Iteration in Dynamic Programming,"  Proc.\ of Allerton Conf.\ on Information Sciences and Systems.

\refer[Ber95] Bertsekas, D.\ P., 1995.\ ``A Counterexample to Temporal Differences
Learning,''  Neural Computation, Vol.\ 7, pp.\ 270-279.

\refer[Ber07] Bertsekas, D.\ P., 2007.\ Dynamic Programming and Optimal Control, 3rd Edition, Vol.\  II, Athena
Scientific, Belmont, MA.

\refer[Ber09] Bertsekas, D.\ P.,  2009.\ ``Projected Equations, Variational Inequalities, and Temporal Difference Methods," Lab.\ for Information and Decision Systems Report LIDS-P-2808, MIT.

\refer[Ber10] Bertsekas, D.\ P.,  2010.\ ``Pathologies of Temporal Difference Methods in Approximate Dynamic Programming," Proc.\ 2010 IEEE Conference on Decision and Control.

\refer[Ber11a] Bertsekas, D.\ P.,  2011.\ Approximate Dynamic Programming, on-line at\hfill\break 
http://web.mit.edu/dimitrib/www/dpchapter.html.

\refer[Ber11b] Bertsekas, D.\ P.,  2011.\ ``Approximate Policy Iteration: A Survey and Some New Methods," J.\ of Control Theory and Applications, Vol.\ 9, pp.\ 310-335.

\refer[Ber11c] Bertsekas, D.\ P.,  2011.\ ``Temporal Difference Methods for General Projected Equations," IEEE Trans.\ on Aut.\ Control, Vol.\ 56, pp.\ 2128-2139.

\refer[Bor08] Borkar, V.\ S., 2008.\ 
Stochastic Approximation: A Dynamical Systems Viewpoint, Cambridge Univ. Press.

\refer[Bor09] Borkar, V.\ S., 2009.\ ``Reinforcement Learning: A Bridge Between Numerical Methods and Monte Carlo," in World Scientific Review, Vol.\ 9, Chapter 4.

\old{
\refer[Boy02]
Boyan, J.\ A., 2002.\
``Technical Update:
Least-Squares Temporal Difference Learning," Machine Learning, Vol.\ 49,
pp.\ 1-15.
}

\refer[BrB96]
Bradtke, S.\ J., and Barto, A.\ G., 1996.\
``Linear Least-Squares Algorithms for
Temporal Difference Learning,''
Machine Learning, Vol.\ 22, pp.\ 33-57.

\refer[CFH07] Chang, H.\ S., Fu, M.\ C., Hu, J., Marcus, S.\ I., 2007.\  Simulation-Based Algorithms for Markov Decision Processes, Springer, N.\ Y.

\refer [CaR11] Canbolat, P.\ G.,  and Rothblum, U.\ G., 2011.\ ``(Approximate) Iterated Successive Approximations Algorithm for Sequential Decision Processes," Technical Report, The Technion - Israel Institute of Technology, May 2011. 

\refer[Cao07] Cao, X.\ R., 2007.\ Stochastic Learning and Optimization: A Sensiti\-vity-Based Approach, Springer, N.\ Y.

\refer[FRF11] Foderaro, G., Raju, V., and Ferrari, S., 2011.\ ``A Model-Based Approximate $\lambda$-Policy Iteration Approach to Online Evasive Path Planning and the Video Game Ms. Pac-Man," J.\ of Control Theory and Applications, Vol.\ 9, pp.\ 391-399.

\refer[Fle84] Fletcher, C.\ A.\ J., 1984.\ Computational Galerkin Methods, Springer-Verlag, N.\ Y.

\refer[Gos03] Gosavi, A., 2003.\ Simulation-Based Optimization
Parametric Optimization Techniques and Reinforcement Learning, Springer-Verlag, N.\ Y.

\ref[Hay08] Haykin, S., 2008.\ Neural Networks and Learning Machines (3rd Edition), Prentice-Hall, Englewood-Cliffs, N.\ J.

\refer[Kra72] Krasnoselskii, M.\ A., et. al, 1972.\ Approximate Solution of Operator Equations, Translated by D.\ Louvish, Wolters-Noordhoff Pub., Groningen.

\refer[LLL08] Lewis, F.\ L., Lendaris, G.\ G., and Liu, D., 2008.\ Special Issue on Adaptive Dynamic Programming and Reinforcement Learning in Feedback Control, IEEE Transactions on Systems, Man, and Cybernetics, Vol.\ 38.

\refer[LaP03] Lagoudakis, M.\ G.,  and Parr, R., 2003.\ ``Least-Squares Policy Iteration," J.\ of Machine Learning Research, Vol.\ 4, pp.\ 1107-1149

\refer[LeV09] Lewis, F.\ L., and Vrabie, D., 2009.\ ``Reinforcement Learning and Adaptive Dynamic Programming for Feedback Control," IEEE Circuits and Systems Magazine, 3rd Q.\ Issue.

\refer[Mey07] Meyn, S., 2007.\ Control Techniques for Complex Networks, Cambridge University Press, N.\ Y.

\old{
\refer [Mun03] Munos, R.. 2003. ``Error Bounds for Approximate Policy Iteration," Proc.\  20th ICML, Washington DC, pp.\ 560Ð567.
}

\refer[NeB03] Nedi\'c, A., and Bertsekas, D.\ P., 2003.\ ``Least Squares Policy Evaluation Algorithms with Linear Function Approximation," Discrete Event Dynamic Systems: Theory and Applications, Vol.\ 13, pp.\ 79-110.

\refer [Pow07] Powell, W.\ B., 2007.\  Approximate Dynamic Programming: Solving the Curses of Dimensionality, Wiley, N.\ Y.

\refer [Put94] Puterman, M.\ L., 1994.\  Markov Decision Processes: Discrete Stochastic Dynamic Programming, J.\ Wiley, N.\ Y.

\refer[Rot79] Rothblum, U.\ G., 1979.\ ``Iterated Successive Approximation for Sequential Decision Processes," in Stochastic Control
and Optimization, by J.\ W.\ B.\ van Overhagen and H.\ C. Tijms (eds), Vrije University, Amsterdam.

\refer[SBP04] Si, J., Barto, A., Powell, W., and Wunsch, D., (Eds.) 2004.\ Learning and Approximate Dynamic Programming, IEEE Press, N.\ Y.

\refer[Sam59] Samuel, A.\ L., 1959.\ ``Some Studies in Machine Learning
Using the Game of Checkers,''  IBM Journal of Research and Development,
pp.\ 210-229.

\refer[Sam67] Samuel, A.\ L., 1967.\ ``Some Studies in Machine Learning
Using the Game of Checkers.\ II -- Recent Progress,'' 
IBM Journal of Research and Development,
pp.\ 601-617.

\refer[Sch10] Scherrer, B., 2010.\ ``Should One Compute the Temporal Difference Fix Point or Minimize the Bellman Residual? The Unified Oblique Projection View," in ICML'10: Proc.\ of the 27th Annual International Conf.\ on Machine Learning.

\refer[Sch11] Scherrer, B., 2011.\ ``Performance Bounds for Lambda Policy Iteration and Application to the Game of Tetris," Report RR-6348, INRIA.

\refer[SuB98] Sutton, R.\  S., and Barto, A.\ G., 1998.\ Reinforcement Learning, MIT
Press, Cambridge, MA.

\refer [Sut88] Sutton, R.\ S., 1988.\ ``Learning to Predict by the Methods of Temporal Differences," Machine Learning, Vol.\ 3, pp.\ 9-44.

\refer[SzL06] Szita, I., and Lorinz, A., 2006.\ ``Learning Tetris Using the Noisy Cross-Entropy Method," Neural Computation, Vol.\ 18, pp.\ 2936-2941.

\refer[Sze10] Szepesvari, C., 2010.\ ``Reinforcement Learning Algorithms for MDPs," Morgan and Claypool Publishers.

\refer[ThS09] Thiery, C., and Scherrer, B., 2009.\ ``Improvements on Learning Tetris with Cross-Entropy," International Computer Games Association Journal, Vol.\ 32, pp.\ 23-33.

\refer[ThS10a] Thiery, C., and Scherrer, B., 2010.\ ``Least-Squares Policy Iteration:
Bias-Variance Trade-off in Control Problems," Proc.\ of 2010 ICML, Haifa, Israel. 

\refer[ThS10b] Thiery, C., and Scherrer, B., 2010.\ ``Performance Bound for Approximate Optimistic Policy Iteration," Technical Report, INRIA.

\refer[TsV97]
Tsitsiklis, J.\ N., and Van Roy, B., 1997.\ 
``An Analysis of Temporal-Difference Learning
with Function Approximation,"
IEEE Transactions on Automatic Control,
Vol.\ 42, pp.\ 674--690.

\ref[WPB09] Wang, M., Polydorides, N., and Bertsekas, D.\ P., 2009.\ ``Approximate Simulation-Based Solution of Large-Scale Least Squares Problems," Lab.\ for
Information and Decision Systems Report LIDS-P-2819, MIT.

\ref[WaB11a] Wang, M., and Bertsekas, D.\ P., 2011.\ ``Stabilization of Simulation-Based Iterative Methods for Singular and Nearly Singular Linear Systems," Lab.\ for Information and Decision Systems Report LIDS-P-2878, MIT.

\ref[WaB11b] Wang, M., and Bertsekas, D.\ P., 2011.\ ``On the Convergence of Iterative Simulation-Based Methods for Singular Linear Systems," Lab.\ for Information and Decision Systems Report LIDS-P-2879, MIT.

\ref[Wer09] Werbos, P.\ J., 2009.\ ``Intelligence in the Brain: A Theory of how it Works and how to Build it," Neural Networks, Vol.\ 22, pp.\ 200-212.

\refer [WhS92] White, D., and Sofge, D., 1992.\ Handbook of Intelligent Control, Van Nostrand Reinhold, N.Y.

\ref[WiB93] Williams, R.\ J., and Baird, L.\ C., 1993.\ ``Analysis of Some Incremental Variants of Policy Iteration: First Steps Toward Understanding Actor-Critic Learning Systems,'' Report NU-CCS-93-11, College of Computer Science, Northeastern University, Boston, MA.

\refer[YuB09] Yu, H., and Bertsekas, D.\ P., 2009.\ ``Convergence Results for Some Temporal Difference Methods Based on Least Squares," IEEE Trans.\ on Aut. Control, Vol. 54, 2009, pp.\ 1515-153.

\refer[YuB10] Yu, H., and Bertsekas, D.\ P., 2010.\ ``Error Bounds for Approximations from Projected Linear Equations," Mathematics of Operations Research, Vol.\ 35, pp.\ 306-329. 

\refer[YuB11] Yu, H., and Bertsekas, D.\ P., 2011.\ ``Q-Learning and Policy Iteration Algorithms 
for Stochastic Shortest Path Problems,"  Lab.\ for Information and Decision Systems Report LIDS-P-2871, MIT.

\refer[Yu10a] Yu, H., 2010.\ ``Least Squares Temporal Difference Methods: An Analysis
Under General Conditions," Technical report C-2010-39, Dept.\ Computer
Science, Univ. of Helsinki.

\refer[Yu10b] Yu, H., 2010.\ ``Convergence of Least Squares Temporal Difference Methods
Under General Conditions," Proc.\ of the 27th ICML, Haifa, Israel.

\end